\documentclass[10pt]{article}
\usepackage[margin=1.2in]{geometry}
\usepackage{graphpap}
\usepackage{calc}
\usepackage{graphicx}
\usepackage{ifthen}
\usepackage{subfigure}
\usepackage{pst-all}
\usepackage{pst-poly}
\usepackage{multido}
\usepackage{dsfont}
\usepackage{array}
\input{random.tex}
\usepackage{hyperref}
\usepackage{xcolor}
 
\usepackage[thinlines]{easytable} 
\usepackage{amsthm} 
\usepackage{amsmath}
\usepackage{amssymb}  
\usepackage{pgfplots}

\newtheorem{theorem}{Theorem}
\newtheorem{definition}{Definition}
\newtheorem{corollary}{Corollary}

\newtheorem{claim}{Claim}
\newtheorem{lemma}[theorem]{Lemma}

\newtheorem{example}{Example}
\newtheorem{prop}[theorem]{Proposition}

\newcommand{\re}{\mathbb{R}}
\newcommand{\na}{\mathbb{N}}

\newcommand{\veps}{\varepsilon}

\newcommand{\beq}{\begin{equation*}}
\newcommand{\eeq}{\end{equation*}}
\newcommand{\head}[1]
{\markright{\hbox to 0pt{\vtop to 0pt{\hbox{}\vskip 3mm \hrule width
\textwidth \vss} \hss}{\sc #1}}}

\begin{document}

\parindent=0pt
\baselineskip18pt
\parskip6pt
\head{Sender-receiver stopping games}

\title{Incentive compatibility in sender-receiver stopping games\thanks{We would like to thank Galit Ashkenazi-Golan, Ga\"{e}tan Fournier, J\'{e}r\^{o}me Renault, Eilon Solan and Bruno Ziliotto for helpful discussions.}}
\date{\today}

\author{Aditya Aradhye \thanks{Dept. of Quantitative Economics, Maastricht University. Email address: a.aradhye@maastrichtuniversity.nl} \hskip6pt
J\'anos Flesch\footnotemark[2] \hskip6pt
Mathias Staudigl\footnotemark[2] \hskip6pt
Dries Vermeulen\footnotemark[2]}

\author{
        Aditya Aradhye
        \thanks{Address of all authors:
        Maastricht University, School of Business and Economics, Dept. of Quantitative Economics,
        P.O.\ Box 616, 6200 MD Maastricht, The Netherlands. E-mail address:
        {\tt a.aradhye@maastrichtuniversity.nl}}
        \and
        J\'{a}nos Flesch
        \thanks{{\tt j.flesch@maastrichtuniversity.nl}}
        \and
        Mathias Staudigl
        \thanks{{\tt m.staudigl@maastrichtuniversity.nl}.}
        \and
        Dries Vermeulen
        \thanks{Corresponding author.{\tt d.vermeulen@maastrichtuniversity.nl}.}
}

\maketitle

\begin{abstract} 
We introduce a model of sender-receiver stopping games, where the state of the world follows an iid--process throughout the game. At each period, the sender observes the current state, and sends a message to the receiver, suggesting either to stop or to continue. The receiver, only seeing the message but not the state, decides either to stop the game, or to continue which takes the game to the next period. The payoff to each player is a function of the state when the receiver quits, with higher states leading to better payoffs. The horizon of the game can be finite or infinite.

We prove existence and uniqueness of responsive (i.e. non-babbling) Perfect Bayesian Equilibrium (PBE) under mild conditions on the game primitives in the case where the players are sufficiently patient. The responsive PBE has a remarkably simple structure, which builds on the identification of an easy-to-implement and compute class of threshold strategies for the sender. With the help of these threshold strategies, we derive simple expressions describing this PBE. It turns out that in this PBE the receiver obediently follows the recommendations of the sender. Hence, surprisingly, the sender alone plays the decisive role, and regardless of the payoff function of the receiver the sender always obtains the best possible payoff for himself.
\end{abstract}

JEL Classification: C73; D82; D83

\emph{Keywords:} {Sender-Receiver games, Stopping games, Bayesian games, Incentive Compatibility

\section{Introduction}                                                    
Information transmission is a fundamental element of economic models. In various settings, a better informed party (sender) is in the position to transmit information to a lesser informed or even uninformed party (receiver). Typically, the action choices of the receiver have an influence on the payoff of the sender, and hence the information transmission has a strategic aspect. In their seminal paper, Crawford and Sobel \cite{CrawfordSobel}, analyze strategic information transmission with a single interaction between the sender and the receiver. Their model and its variations \cite{CrawfordSobel,GreenStokey} have a wide range of applications, notably in economics, computer science, political science but also in biology and philosophy \cite{CShumanagent,SignalsEvolution,dynamics}. Recently, a few models have been introduced in which the information transmission takes place in a dynamic setting, see for instance \cite{Renault13} and \cite{Golosov2014}. In these models, the sender-receiver game is played repeatedly either on a finite or an infinite horizon and the payoff for the sender and the receiver is the total discounted sum of the stage payoffs. The key focus in these papers is the characterization of the set of equilibrium payoffs where, in the spirit of the folk theorem, the players are sufficiently patient.

This paper sets the stage for a different line of research, which we regard as an important conceptual contribution of our work. Specifically, this paper introduces a model of \emph{sender-receiver stopping games}, which combines features from dynamic sender-receiver games and stopping games (for a survey on the latter, see \cite{stoppingsurvey}). In these games the strategic information transmission takes place repeatedly until the receiver decides to stop the interaction. More precisely, we deal in this paper with sender-receiver stopping games of finite as well as infinite horizon. In the finite horizon, the receiver is forced to stop the game before a pre-defined terminal period has been reached. In the infinite horizon the game may be played for unlimited number of periods. The timing of the game is as follows: In each period nature draws a state of the world, which is only revealed to the sender. After observing the state of the world, the sender sends one out of two messages to the receiver. This message is interpreted as a suggestion either to stop the game or to continue. Now the receiver has to take a decision. After seeing the message, but without knowing the state, the receiver has two options: he can decide to stop the game, or he can decide to continue to the next period. The payoff to each player is a function of the state at which the receiver stops, and these payoffs are either discounted or undiscounted. The setting could be thought as an investor (the receiver) who must make an irreversible financial decision, without having exact information about the market situation, but using the advice of an expert (the sender).

We assume that rewards are positively correlated with the state of nature, i.e. higher states lead to better payoffs for both players. Thus, both player have identical ordinal preferences over realizations of the state of nature. Yet, as we impose no further restrictions on the payoffs, the cardinal assignment of values of the two payoff functions can be very different. As a consequence, a certain state may be very appealing to one player, but not so much to the other, creating an interesting strategic tension between the parties. This paper investigates to what extent the cardinal differences may hamper coordination between the players. 

The main solution concept that we use to analyze these games is \emph{Perfect Bayesian Equilibrium} (PBE), and we identify a class of PBEs which are appealing to the players in terms of payoffs, and moreover easy to compute and implement.  

\subsection{Our contribution}
We are interested in PBEs in which the receiver plays a responsive strategy.  We call a receiver's strategy responsive if his mixed stage game action is different for different messages sent by the sender. This is just a condition which excludes those PBEs in which the receiver's strategy is babbling, which are fairly uninteresting from a strategic perspective. \footnote{Regardless, it is fairly easy to study all the PBEs which are not responsive by simply adding babbling periods in any responsive PBE.}

Therefore, the crucial question is how the sender should use his additional information to manipulate the choices of the receiver given the current state, and to what extent the receiver can trust the sender's recommendations and be obedient. In particular, we investigate under what conditions the incentives of the sender and the receiver match. 

These concerns are captured by the notion of a \emph{regular strategy profile}. In a regular strategy profile, the receiver simply follows the sender's recommendations, whereas the sender sends a sincere message, given the realization of the state, whether or not he would like the receiver to terminate the game at this period. Since the receiver is obedient, he is not playing an active role in such a strategy profile. The sender's sincere strategy is a threshold strategy that sends the message ``continue" if the current state is below the threshold and sends the message ``quit'' otherwise. In other words, the sender's strategy is the optimal solution of the one player maximization problem in which the decision to continue or to quit is delegated to the sender. This means that in the regular strategy profile the sender obtains the best possible payoff for himself. Indeed, this outcome is Pareto optimal. The game we study admits an (essentially) unique\footnote{By (essentially) unique, we mean that in any two regular strategy profiles, the actions of the players differ at only measure zero sets and thus induce the same expected payoffs for the players.} regular strategy profile. 

In finite horizon, we show that there is no responsive PBE other than the regular strategy profile. Even the regular strategy profile may fail to exist in certain games if the discount factor is small. More precisely, our findings are as follows:
\begin{itemize}
\item[(i)] The regular strategy profile is the unique responsive PBE if the discount factor is sufficiently high or if the payoffs are undiscounted. 
\item[(ii)] There are games that have no responsive PBE if the discount factor is small enough.
\end{itemize}
For the infinite horizon, we focus on strategy profiles where the expected payoffs after any period do not depend on the history. We show that within this class of strategy profiles, the regular strategy profile is the unique responsive PBE provided the discount factor is sufficiently high. More precisely, our findings are as follows:
\begin{itemize}
\item[(iii)] The regular strategy profile is the unique responsive PBE if the discount factor is sufficiently high. In this setting, the regular strategy profile is stationary.
\item[(iv)]  There are games that have no responsive PBE if the discount factor is small enough.
\end{itemize}
In the extreme case of infinite horizon where the payoffs are undiscounted, there does not exist any responsive PBE. This is due to the fact that the sender does not have a best response when the receiver is obeying, and hence, it should not be interpreted as a breakdown of communication.

\subsection{Related literature}

Crawford and Sobel \cite{CrawfordSobel} introduced a model of strategic information transmission. The model in which  the sender and the receiver interact only once is studied extensively, see \cite{CrawfordSobel,GreenStokey}. Recently a lot of work has been focused on the dynamic extension, where the strategic interaction takes place repeatedly either for finite or infinite number of periods, see \cite{Renault13, Golosov2014, artconv, longcheap}. \cite{Renault13} assume that the sequence of states follows an irreducible Markov chain. They characterize the limit set of equilibrium payoffs, as players become very patient. \cite{Golosov2014} study finite horizon games, and show that, under certain conditions, full information revelation is possible and conditioning future information release on past actions improves incentives for information revelation. 

Our paper relates to the large and growing literature of Bayesian persuasion, see \cite{kamenica2011bayesian,ely2017beeps,renault2017optimal,honryo2011dynamic}. In these settings, the informed advisor (i.e. the sender) decides how much information to share with a less informed agent (i.e. the receiver) so as to influence his decision. \cite{renault2017optimal} show that in many cases, the optimal greedy disclosure policy for the sender exists, which at each stage, minimizes the amount of information being disclosed in that stage under the constraint that it maximizes the current payoff of the sender. 

This paper also relates to the classical contributions on communication in games (see e.g. \cite{forges1986approach,myerson1986multistage}). In \cite{skyrms2010flow,blume1998experimental}, the evolution of information flow is studied in the setting of strategic communication. Our model is also linked to topics in computer science such as automated advice provision. For instance, \cite{CShumanagent} use sender-receiver games to model interaction between computers and humans.

Most of the previous work has focused on the sender-receiver games with fixed duration of time. Compared to these earlier papers, one of the main novelties of our model is that the game is a stopping game. The receiver has the license to stop the game at any period of time. Some of the techniques we use (for example, backward induction) are also similar to the ones in the literature on stopping games, see \cite{stoppingsurvey,valuestopping}. 

In the responsive PBE, the receiver complies with the sender and the sender tries to maximize his expected payoff without knowing the future states. Hence, our setting in a wider sense, is a variant of the secretary problem, see \cite{ferguson1989solved}.

The structure of the paper is as follows. In section \ref{sec-model} we introduce the model and in the section \ref{sec:reg}, we discuss the regular strategy profile. In sections \ref{sec:fin} and \ref{sec:infin} we state our main results for the finite horizon and the infinite horizon respectively. In section \ref{sec:ex}, we give illustrative examples and in section \ref{sec:conclude}, we have concluding remarks. In section \ref{proof}, we provide the proofs of the main theorems.

\section{The model}\label{sec-model}

In this section we describe the model of sender-receiver stopping games and the solution concept of Perfect Bayesian Equilibrium.

\subsection{The game}
We study sender-receiver stopping games. These are dynamic games played by two players, the sender and the receiver, either with finite or infinite horizon. 

An infinite horizon sender-receiver stopping game is played at periods in $\na = \{1, 2, \ldots \}$.
At period $t\in\na$, play is as follows. First, a state of the world $\theta^t$ is drawn uniformly from the unit interval $I = [0,1]$,
independently of the earlier realizations $\theta^1, \ldots, \theta^{t-1}$. The sender learns $\theta^t$, while the receiver only knows the distribution of $\theta^t$.
Next, the sender chooses a message $m^t \in \{m_c, m_q\}$ and sends it to the receiver.
The message $m^t = m_c$ is interpreted as a suggestion for the receiver to continue at this period $t$ and the message $m^t = m_q$ as a suggestion to quit.
On seeing the message, the receiver chooses an action $a^t \in \{a_c, a_q\}$, where $a_c$ stands for continue and $a_q$ stands for quit.
If the receiver quits then the game ends at period $t$, whereas if the receiver continues then the game proceeds to period $t+1$.
If the game ends at period $t$, then the sender receives the payoff $f^t(\theta^t)$ and the receiver receives the payoff $g^t(\theta^t)$.
Here, $f^t$ and $g^t$ are two continuous and strictly increasing functions from $I$ to $\re_+$.
If the receiver never quits, both players receive payoff zero, with $f^t(0) = g^t(0) = 0$. For each player, the outcome in which the receiver continues forever is worst. The payoff if the receiver continues forever is zero for each player.

If there are functions $f$ and $g$ such that $f^t = f$ and $g^t = g$ for all periods $t\in\na$, we say that the game has period-independent payoffs.
If there are functions $f$ and $g$ and $\delta \in(0,1)$ such that $f^t = \delta^{t-1} \cdot f$ and $g^t = \delta^{t-1} \cdot g$ for all periods $t\in\na$,
we say that the game has discounted payoffs with discount factor $\delta$. In such cases, the functions $f$ and $g$ are called \emph{characteristic functions} of the game. 
We assume throughout the paper that payoffs are either period-independent or discounted.

The model description for sender-receiver stopping games on a finite horizon $T\in\na$ is almost identical to the one above. The only modification is that if period $T$ is reached then the game terminates at the end of period $T$. Now we proceed with the description of sender-receiver stopping games on the infinite horizon, without explicitly mentioning the changes for the finite horizon.

\subsection{Strategies and expected payoffs}\label{str+exp}

\textbf{Histories.} For the sender, a history at period $t$ is a sequence
$h_s^t=(\theta^1, m^1, \ldots, \theta^{t-1}, m^{t-1})$
of past states and messages sent by the sender.
By $H_s^t = (I \times M)^{t-1}$ we denote the set of histories for the sender at period $t$.
Given the usual topology on $I$, we endow $H_s^t$ with the product Borel sigma-algebra.

Since the receiver does not observe the realization of the states, a history of the receiver at period $t$ is a sequence
$h_r^t = (m^1,\ldots, m^{t-1})$
of past messages sent by the sender. By $H_r^t = M^{t-1}$ we denote the set of histories for the receiver at period $t$.
Note that $H_r^t$ is a finite set. 
\vskip6pt

\textbf{Strategies.} A strategy $\sigma = (\sigma^t)_{t=1}^\infty$ for the sender is a sequence of measurable functions $\sigma^t \colon H_s^t \times I \rightarrow [0,1]$.
The interpretation is that, at each period $t$, given the history $h_s^t$ and the state $\theta^t$,
the strategy $\sigma^t$ places probability $\sigma^t(h_s^t,\theta^t)$ on the message $m_c$.

A strategy $\tau = (\tau^t)_{t=1}^\infty$ for the receiver is a sequence of functions $\tau^t \colon H_r^t \times M \rightarrow [0,1]$.
We do not need any measurability conditions for $\tau^t$ as the domain of $\tau^t$ is finite.
The interpretation is that, at each period $t$, given the history $h_r^t$ and the message $m^t$,
the strategy $\tau^t$ places probability $\tau^t(h_r^t, m^t)$ on the action $a_c$.

For the case when the game has a finite horizon $T$, for simplicity we require that at period $T$, regardless the history, the sender's strategy has to send the message $m_q$ and the receiver's strategy has to play the action $a_q$.

In this model, we focus on the \textit{responsive} strategies of the receiver. A strategy $\tau$ of the receiver is called responsive if, for each period $t$ (with $t<T$ if the game has finite horizon $T$) and history $h_r^t$,
we have $\tau^t(h_r^t, m_c) > \tau^t(h_r^t, m_q)$.
This is saying that, upon receiving the message $m_c$, the receiver chooses action $a_c$ with higher probability than upon receiving $m_q$.
\footnote{The reason to restrict our attention to responsive strategies is to avoid PBEs in `babbling' strategies, which are fairly uninteresting from a game theory perspective.}
\vskip6pt

\subsection{Perfect Bayesian equilibrium}
In this section we introduce the solution concept we use to analyse the sender-receiver stopping games defined above.

Consider a strategy profile $(\sigma,\tau)$. The expected payoffs of the sender and the receiver are denoted by $U_s(\sigma,\tau)$ and $U_r(\sigma,\tau)$ respectively. If the receiver has not quit until some period $t$, and the histories are $h^t_s$ and $h^t_r$ respectively, then the continuation expected payoffs from period $t$ onward are denoted by $U^t_s(\sigma,\tau)(h^t_s)$ and $U^t_r(\sigma,\tau)(h^t_r)$ respectively. For details on the definitions of these notations, refer to the Appendix B.

We say that, at period $t$, the expected payoff $U^t_s(\sigma, \tau)$ for the sender is history independent if for every $h_s^t, \overline{h}_s^{t} \in H_s^t$ it holds that
$U^t_s(\sigma,\tau)(h_s^t) = U^t_s(\sigma,\tau)(\overline{h}_s^{t})$. Note that history independence of $U^t_s(\sigma, \tau)$ is equivalent to saying that
the function $U^t_s(\sigma, \tau)$ is constant. In that case, with slight abuse of notation, we identify the function with the (constant) value of that function,
and act as if $U^t_s(\sigma, \tau)$ is a real number instead of a function. A similar observation holds for the expected payoff $U^t_r(\sigma, \tau)$ for the receiver.

\begin{definition}\label{PBE}
A strategy profile $(\sigma, \tau)$ is called a Perfect Bayesian Equilibrium (PBE) if for every period $t$, and every history $h_s^t$,
we have $U_s^t(\sigma, \tau)(h_s^t) \ge U_s^t(\sigma', \tau)(h_s^t)$ for every  strategy $\sigma'$ of the sender,
and for every $h_r^t$, $U_r^t(\sigma, \tau)(h_r^t) \ge U_r^t(\sigma, \tau')(h_r^t)$ for every strategy $\tau'$ of the receiver. A PBE is called responsive if the receiver's strategy is responsive.
\footnote{\rm 
PBE is a refinement of Bayesian Nash Equilibrium (BNE). Intuitively, it requires that the strategy profile induces a BNE after any history.}
\end{definition}

Notice that in the definition of PBE we do not explicitly talk about beliefs of the players on the realized history consisting of the past states, messages and actions. Since the sender's history contains all this information, he is fully informed and he knows the history of the receiver. On the other hand, the receiver is not informed of the past or current states. Based on his own history and the strategy profile $(\sigma,\tau)$, he has a natural belief on the possible histories of the sender, which is compatible with Bayesian updating. For details, we refer to Appendix A. 
\vskip6pt

We will regularly make use of the fact that the well-known one-shot deviation principle holds in our games whenever the game has a finite horizon (regardless whether the payoffs are period-independent or discounted) or the game has infinite horizon and the payoffs are discounted. More precisely, in these settings, a strategy profile $(\sigma, \tau)$ is a PBE if and only if (1) for every history $h_s^t$ of the sender, we have $U_s^t(\sigma, \tau)(h_s^t) \ge U_s^t(\sigma', \tau)(h_s^t)$ for every $\sigma'$ that is a one-shot deviation from $\sigma$ at $h_s^t$, and (2) similarly for the receiver. Here, for two strategies $\sigma$ and $\sigma'$ and a history $h_s^t$ of the sender, $\sigma'$ is called a one-shot deviation from $\sigma$ at history $h_s^t$ if $\sigma'(h)=\sigma(h)$ for every history $h\neq h_s^t$ of the sender. One-shot deviations are defined similarly for the receiver.

\subsection{Terminology for strategies}\label{spec-str}

A strategy $\sigma$ for the sender is called \emph{pure} if, for each period $t$, history $h_s^t$ and state $\theta^t$,
either $\sigma^t(h_s^t,\theta^t) = 1$ or $\sigma^t(h_s^t,\theta^t) = 0$. Pure strategies for the receiver are defined in a similar fashion.
\vskip6pt

A strategy $\sigma$ for the sender is said to have a \emph{threshold at period $t$}, if there exists a threshold $\beta^t \in [0,1]$ such that 
$$
\sigma^t(\theta^t) =
\begin{cases}
1 & \hbox{if } \theta^t \in [0,\beta^t) \cr
0 & \hbox{if } \theta^t \in (\beta^t,1]. 
\end{cases}
$$
We do not specify what the strategy recommends when the state is exactly equal to the threshold, for the sake of flexible exposition of our results. In any case, this occurs with probability zero only. A strategy $\sigma$ for the sender is called a \emph{threshold strategy} if it has a threshold at each period $t$. 
A threshold strategy $\sigma$ is called \emph{stationary} if $\beta^s = \beta^t$ for all periods $s$ and $t$.
\vskip6pt

A strategy profile $(\sigma,\tau)$ is called \emph{essentially Markov} if $U^t_s(\sigma,\tau)$ and $U^t_r(\sigma,\tau)$ are history independent. So the history at period $t$ does not influence the continuation payoffs from period $t$ onward, although it still may influence the continuation strategies.

\section{The regular strategy profile}\label{sec:reg}

The regular strategy profile plays a central role in our paper. A strategy profile $(\sigma,\tau)$ is called \emph{regular} if
$\tau$ is the obedient strategy, and $\sigma$ is sincere against $\tau$.
The \emph{obedient strategy} $\tau$ for the receiver is defined, for each period $t$ and each history $h_r^t$, by
$\tau^{t}(h_r^t, m_c) = 1$ and $\tau^{t}(h_r^t, m_q) = 0$. The obedient strategy is pure, and responsive.
\vskip6pt

For a given strategy $\tau$ of the receiver,
a threshold strategy $\sigma$ of the sender is called \emph{sincere against $\tau$ at period $t$} (with $t<T$ if the game has finite horizon $T$) if
\begin{itemize}
\item[{[1]}] $U^{t+1}_s(\sigma, \tau)$ is history independent, and
\item[{[2]}] the strategy $\sigma$ has the threshold $\alpha^t$ at period $t$ where $\alpha^t$ is the solution to the equation
$f^t(\alpha^t) = U_s^{t+1}(\sigma,\tau)$.
\end{itemize}
A threshold strategy $\sigma$ is called \emph{sincere against} $\tau$ if it is sincere against $\tau$ at each period $t$ (with $t<T$ if the game has finite horizon $T$).
\vskip6pt

Notice that indeed the equation in condition [2] has a unique solution due to monotonicity of $f^t$,
and the fact that, since payoffs are either period-independent or discounted, $U_s^{t+1}(\sigma,\tau) \le f^t(1)$.
\vskip6pt

In a regular strategy profile, the sender sends a sincere message
whether or not he would like the receiver to terminate the game at this period. Next paragraph provides intuitive explanation on why the condition [2] achieves this.

Assume that $(\sigma,\tau)$ is a regular strategy profile such that $\sigma$ has threshold $\alpha^t$ at period $t$. If $\theta^t < \alpha^t$, then $f^t(\theta^t) < f^t(\alpha^t) = U_s^{t+1}(\sigma,\tau)$. In this case the sender would like the receiver to continue the game as the expected continuation payoff is higher than the expected payoff if the receiver quits. Indeed, the strategy $\sigma$ recommends the message $m_c$ as $\theta^t < \alpha^t$ and $\alpha^t$ is the threshold. Similarly, if $\theta^t > \alpha^t$, then $f^t(\theta^t) > f^t(\alpha^t) = U_s^{t+1}(\sigma,\tau)$ and $\sigma$ recommends the message $m_q$.

We argue in both the finite and the infinite horizon model that the regular strategy profile is (essentially) unique, and entirely computable. To explain this, we define the auxiliary function $H \colon I \rightarrow I$ by
$$
H(x) = f^{-1} \Big( \delta \cdot \Big[ x \cdot f(x) + \int_x^1 f(\theta) d\theta \Big] \Big).
$$
\vskip6pt 

\subsection{Finite horizon}

Assume that the game has finite horizon $T$. Define the numbers $\beta^1, \ldots, \beta^T$ as follows\footnote{
Note that $\beta^t$ depends on the horizon $T$. When needed we write $\beta^t(T)$ instead of $\beta^t$.}.
First, $\beta^T = 0$. Then, using a backwards iteration, $\beta^t = H(\beta^{t+1})$ for all $t=T-1,\ldots,1$. We have (cf. Lemma \ref{lem:fin}.1)
\[1 > \beta^1>\beta^2>\cdots > \beta^{T}=0.\]

\begin{prop} \label{sinc}
Assume that the game has finite horizon $T$. Then the thresholds for the sender's strategy in the regular strategy profile are given by $\beta^1, \ldots, \beta^T$ respectively at periods $1,\ldots,T$.
\end{prop}

Proof. \quad
Suppose that the players use a regular strategy profile $(\sigma,\tau)$. Assume that $\sigma$ has a threshold $\alpha^t$ at each period $t$. Then, at each period $t<T$, with probability $\alpha^t$ we have $\theta^t < \alpha^t$ and the receiver continues. In this case the sender gets the expected continuation payoff $U_s^{t+1}(\sigma,\tau)$. Similarly, with probability $1-\alpha^t$ we have $\theta^t > \alpha^t$ and the receiver quits. In this case the sender gets the expected continuation payoff $\frac{1}{1-\alpha^t}\int_{\alpha^t}^1 f^t (\theta) d\theta$. As $\sigma$ is sincere against $\tau$, we have $f^t(\alpha^t) = U_s^{t+1}(\sigma,\tau)$ for $t=1,\ldots,T-1$. This yields the recursive equation 
$$
f^{t-1}(\alpha^{t-1}) \,=\, U_s^{t}(\sigma,\tau) \,=\, \alpha^t \cdot \,U_s^{t+1}(\sigma,\tau) + \int_{\alpha^t}^1 f^t(\theta) d\theta \;=\; \alpha^t \cdot f^t(\alpha^t) + \int_{\alpha^t}^1 f^t(\theta) d\theta.
$$ 
Using the fact that $f^t = \delta^{t-1} \cdot f$ in every period $t$, this can be rewritten to 
\[ \alpha^{t-1} = f^{-1}\Big(\delta \cdot\Big[ \alpha^t \cdot f(\alpha^t) + \int_{\alpha^t}^1 f(\theta) d\theta\Big]\Big).\]
The last equation shows that $\,\alpha^{t-1}=H(\alpha^t)\,$ for $t = 2,\dots,T$. As the sender must send the message $m_q$ at period $T$ irrespective of the state,
we have $\alpha^T = 0$. So, we have $\alpha^T = \beta^T$ and inductively, $\,\alpha^{t-1}=H(\alpha^t) = H(\beta^t) = \beta^{t-1}\,$ for $t = T,\dots,2$.
\hfill \qed
\vskip6pt

The numbers $\beta^1,\dots,\beta^T$ can be computed recursively. Hence the regular strategy profile is entirely computable. It is worth noting that the computation of the regular strategy profile only considers the sender's payoff function; the receiver's payoff function does not play any role at all. 

\subsection{Infinite horizon}

The function $H$ has a unique fixed point, which is denoted by $\beta$ (ref. Lemma \ref{lem:func}.2).

\begin{prop} \label{sinc2}
Assume that the game has infinite horizon. 
Then there is a unique regular strategy profile. Moreover, this profile                     is stationary with threshold $\beta$.
\end{prop}

Proof. \quad
Let $(\sigma, \tau)$ be a regular strategy profile. First note that the thresholds $\alpha^t$ of the regular strategy profile satisfy the recursive formula
$\alpha^t = H(\alpha^{t+1})$ for all $t$.
\vskip6pt

We first show that, for each period $t \in \mathbb{N}$ we have $\alpha^t \le \beta$.
Assume by way of contradiction that for some $t \in \mathbb{N}$, $\alpha^t > \beta$. We have $\alpha^t = H(\alpha^{t+1})$. By Lemma \ref{lem:func}.3,
$H\big( \alpha^t \big) < \alpha^t = H(\alpha^{t+1})$. This implies $\alpha^t < \alpha^{t+1}$. In particular, $\alpha^{t+1} > \beta$,
and we can conclude that the sequence $(\alpha^{t'})_{t'=t}^{\infty}$ is strictly increasing. 
\vskip6pt

By definition, $\alpha^t \le 1$ for each $t \in \mathbb{N}$. Moreover, $\alpha^{t'} = H\big(\alpha^{t'+1}\big) \le H(1) < 1$ for each $t'>t$.
Hence, the sequence converges, say to $r < 1$. Write
$$
z = \min \{ x - H(x) \mid x \in [\alpha^t, 1] \}.
$$
By Lemma \ref{lem:func}.3, $H(x) < x$ for $x \ge \alpha^t > \beta$. So, by continuity of $H$, we know that $z > 0$.
As the sequence $(\alpha^{t'})_{t'=t}^\infty$ converges to $r$, for $\epsilon = \frac{z}{2}$, there exists $N>0$ such that $\alpha^{t'} \in [r - \epsilon, r]$ for all $t'>N$.
Then for any $t' > N$, it holds that
$$
\alpha^{t'+1} \ge H\big(\alpha^{t'+1}\big) + z = \alpha^{t'} + z \ge r - \epsilon + z = r + \frac{z}{2}.
$$
This contradicts the fact that $\alpha^{t'+1} > \beta$. Hence, $\alpha^t \le \beta$ for each $t \in \mathbb{N}$.
\vskip6pt

We show that $\alpha^t = \beta$ for each $t \in \mathbb{N}$.
Assume that $\alpha^t < \beta$ for some $t$. As $\alpha^t = H\big(\alpha^{t+1}\big)$, by Lemma \ref{lem:func}.3, $\alpha^{t+1} < \beta$.
So, the sequence $(\beta^t)_{t=1}^\infty$ is decreasing and bounded below by $0$. Hence, the sequence converges, say to $r \ge 0$.
\vskip6pt

Take $\veps = \frac{1}{2} (H(r) - r)$. As before, we can conclude that $\veps > 0$. 
As the sequence $(\alpha^t)_{t=1}^\infty$ converges to $r$, we can find $N > 0$ such that $\alpha^N - r < \epsilon$.
Also, as the sequence is decreasing, $r < \alpha^{N+1}$ and hence $H(r) < H(\alpha^{N+1})$. So, we have
$$
H(r) - r < H(\alpha^{N+1}) - r =  \alpha^N - r < \epsilon = \frac{1}{2} (H(r) - r)
$$
This is a contradiction. Hence, $\alpha^t \ge \beta$. By the previous argument, it follows that $\alpha^t = \beta$.
\hfill \qed

The fixed point $\beta$ can be computed by solving $H(x)=x$. Hence the regular strategy profile is entirely computable. As in the finite horizon, the computation of the regular strategy profile only considers the sender's payoff function; the receiver's payoff function does not play any role at all. 

By Lemma \ref{lem:fin}.3. we have $\beta^t(T) \rightarrow \beta$ as $T \rightarrow \infty$ for each $t$. So the following corollary is immediate consequence of Propositions \ref{sinc} and \ref{sinc2}.

\begin{corollary}\label{regconv}
Let $(\sigma_T,\tau_T)$ be the regular strategy profile in the game with finite horizon $T$ and $(\sigma,\tau)$ be the regular strategy profile in the game with infinite horizon. 
Then the sequence $(\sigma_T,\tau_T)_{T=1}^{\infty}$ converges to $(\sigma,\tau)$ as $T\rightarrow \infty$, when the payoffs are discounted or undiscounted. 
\end{corollary}

\section{Existence and unicity of PBE, finite horizon}\label{sec:fin}

In this section we consider the case where the game has some finite horizon $T$. We provide a existence and unicity result for the PBE of the game.
The result shows that, for any finite horizon, if the payoffs are period-independent or they are discounted with a large discount factor, then the regular strategy profile is the unique responsive PBE. This means that the strategy profile in which the receiver is obedient and the sender is being sincere against this strategy of the receiver
is the only PBE that is responsive. We also show that existence of PBE may fail for small discount factors.
\vskip6pt

Define the function $V \colon [0,1] \rightarrow \re$ by
$$
V(x) = \frac{1}{x} \cdot \int_0^x g(\theta) d\theta
$$
and $V(0) = 0$. The amount $V(x)$ is the expected payoff for the receiver if he quits, conditional on the state being in $[0,x]$.
For $T \in \mathbb{N}$, let $D^T$ be the smallest number \footnote{The number $\beta^1(T)$ itself depends on the discount factor $\delta$. Hence, $D^T$ does not have a simple closed formula. It turns out that $D^T<1$.} in $[0,1]$ such that $\delta \cdot V(1) \ge V(\beta^1(T))$ for every $\delta \in [D^T,1]$.

\begin{theorem} \label{thm:char}
Consider a sender-receiver stopping game with finite horizon $T$. Let the payoffs be either period-independent or discounted with discount factor $\delta \ge D^T$.
Then, the regular strategy profile is the unique responsive PBE.
\end{theorem}
  
Let $D$ be the smallest number in $[0,1]$ such that $\delta \cdot V(1) \ge V(\beta)$ for every $\delta \in [D,1]$. If $f$ is Lipschitz at 1 then $D<1$, as is shown by Lemma \ref{lem:infin}.2.  
Notice that if $\delta \ge D$, then the existence and unicity holds regardless the horizon $T$ of the game.

\begin{theorem}\label{exist}
Suppose that $0 < \delta < D^2$. Then, for any $T \ge 2$, the sender-receiver game with finite horizon $T$ does not admit a responsive PBE.
\end{theorem}

In these theorems, whether the regular strategy profile is a PBE (ref Theorem \ref{thm:char}) or not (ref Theorem \ref{exist}) depends on whether the receiver is patient enough, that is, whether the discount factor is sufficiently high. In particular, it depends on whether he is willing to obey the sender and continue when the state is small. How high the discount factor should be, depends on the curvature of the receiver's payoff function and the thresholds set by the sender.

In the following example, if the discount factor is small then receiver is not inclined to obey the sender at period $T-1$ when the state is small. Hence, the regular strategy profile is not a PBE.

\begin{example} \rm
Consider the game with finite horizon $T$ in which the payoff functions are $\delta$ discounted with $f(\theta) = \theta^2$ and $g(\theta) = \theta$. Then
$$
H(x) = \sqrt{\frac{\delta \cdot (1 + 2 x^3)}{3}}.
$$
If the message sent by the sender at period $T-1$ is $m_c$, then the state is in the interval $[0,\beta^{T-1}]$. So, the payoff for the receiver on quitting is $\delta^{T-2} \cdot V(\beta^{T-1})$ and the payoff on continuing is $\delta^{T-1} \cdot V(1)$. Hence, the receiver prefers to continue if  $\delta \cdot V(1) \ge  V(\beta^{T-1})$. We have $\beta^{T-1} = \sqrt{\frac{\delta}{3}}$. Further, $V(x) = \frac{x}{2}$.
Thus, the inequality $\delta \cdot V(1) \ge  V(\beta^{T-1})$ is valid if and only if $\delta \ge \frac{1}{3}$. 
So, we get $D^2 = \frac{1}{3} > 0$. Hence, if $\delta < \frac{1}{3}$, the regular strategy profile is not a PBE. By Theorem \ref{exist}, if $\delta < \frac{1}{3}$ then the game does not admit a responsive PBE.
\hfill \qed 
\end{example}

\section{Existence and unicity of PBE, infinite horizon} \label{sec:infin}

Now we consider sender-receiver games with infinite horizon. Recall that if the receiver never quits then both players get payoff zero. The payoffs are either period-independent or they are discounted. For the discounted case with sufficiently large discount factors, we prove the existence of a unique responsive PBE, which turns out to be stationary. Then, we show that if the payoffs are period-independent then a PBE fails to exist. 

\begin{theorem}\label{stat-thm}
Consider a sender-receiver stopping game with infinite horizon in which the payoffs are discounted with discount factor $\delta \geq D$.
Then the regular strategy profile is the unique responsive PBE among the essentially Markov strategy profiles.  
\end{theorem}
 
Note that, by Proposition \ref{sinc2}, the regular strategy profile is stationary, with threshold $\beta$. Now we turn to period-independent payoffs.

\begin{theorem}\label{nonex} 
Consider a sender-receiver stopping game with infinite horizon in which the payoffs are period-independent.
Then, there exists no responsive PBE within essentially Markov strategy profiles.
\end{theorem}

The result of theorem \ref{nonex} is driven by the following observation. At period 1, it has probability 1 that a state strictly less than 1 is realized. Since the horizon of the game is infinite, the sender knows that if both players wait sufficiently long, then a strictly better state will be realized later on. Hence, he would like the game to continue. Since this argument holds for each period, the sender is never satisfied. In fact, the players can get payoffs arbitrarily close to $f(1)$ and $g(1)$, but with probability 1 they can not get them exactly. Note that in the regular strategy profile, the threshold in the sender's strategy would be 1 at each period.

The following Corollary is the immediate consequence of Corollary \ref{regconv} and Theorems \ref{thm:char}, \ref{stat-thm} and \ref{nonex}.

\begin{corollary}
If the payoffs are discounted ($\delta<1$), the sequence of PBEs in the finite horizon games converges to a PBE in the infinite horizon game. If the payoffs are undiscounted ($\delta=1$), the sequence of PBE in the finite horizon games converges to the regular strategy profile in the infinite horizon game which is not a PBE.  
\end{corollary}

\section{Examples}\label{sec:ex}
In this section we illustrate our results with the help of the examples. We consider a sender-receiver stopping game in which the payoffs are either period-independent $(\delta=1)$ or discounted with discount factor $\delta\in(0,1)$. As discussed before, the regular strategy profile is determined solely by the payoffs of the sender. 

\begin{example} \rm
$f(x) = x^2$ and $g(x) = x$.
The function $H$ can be calculated as $H(x) = \sqrt{\delta \cdot \frac{1+2x^3}{3}}$. 

We first consider the setting in which the game has a finite horizon $T$. By definition, we have $\beta^T(T) = 0$ and for each $t<T$, $\beta^t(T) = H(\beta^{t+1}(T))$. By Proposition \ref{sinc}, the game has a unique regular strategy profile. In this profile, the receiver plays the obeying strategy and the sender plays the sincere strategy (against the receiver's strategy) with threshold $\beta^t(T)$ at period $t$. 

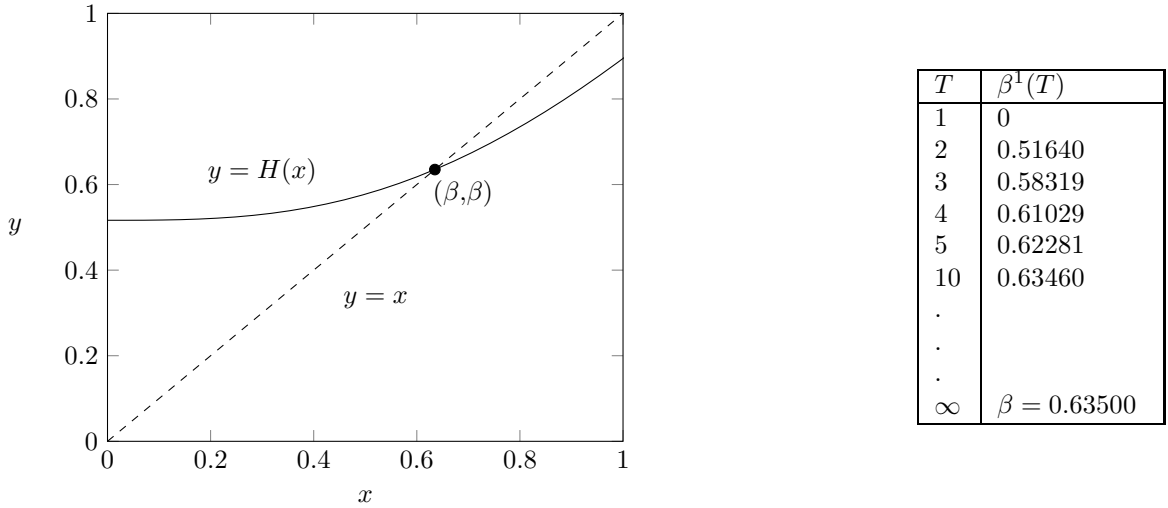
\begin{figure}[!ht]
\centering
    \begin{tikzpicture}[baseline=2.5cm]
  \begin{axis}[xlabel=$x$,ylabel=$y$,ylabel style={rotate=-90}, xmin = 0, xmax = 1, ymin = 0, ymax = 1] 
    \addplot [black, domain=0:3.5,samples=100] {(0.8*(1+2*x^3)/3)^(0.5)};
    \addplot [black, dashed, domain=0:3.5,samples=100] {x};
    \node at (30,63) {$y=H(x)$};
    \node at (52,33) {$y=x$};
    \filldraw[black] (63.5,63.5) circle (2pt);
    \node at (69,58) {$(\beta$,$\beta$)};
  \end{axis}
\end{tikzpicture} \hfill
    \begin{tabular}{ |m{0.4cm} | m{2cm}|} 
    \hline
    $T$ & $\beta^1(T)$\\ 
    \hline
    $1$ & $0$      \\  
    $2$ & 0.51640  \\ 
    $3$ & 0.58319  \\ 
    $4$ & 0.61029      \\  
    $5$ & 0.62281 \\ 
    $10$ & 0.63460  \\ 
    .    &          \\
    .    &          \\
    .    &          \\    
    $\infty$ & $\beta = 0.63500$  \\

 \hline
\end{tabular}
\caption{$f(x)=x^2$, $\delta=0.8$}
\end{figure}

Now as $g(x) = x$, the function $V$ can be calculated as $V(x) = \frac{x}{2}$. Recall from Section \ref{sec:fin} that the bound $D^T$ is defined as the smallest number such that $\delta \cdot V(1) \ge V(\beta^1(T))$ for $\delta \in [D^T,1]$. The inequality simplifies to $\delta \ge \beta^1(T)$. 

If the horizon $T=3$, then $\beta^3(3) = 0$, $\beta^2(3) = \sqrt{\frac{\delta}{3}}$ and $\beta^1(3) = \sqrt{\frac{3\sqrt{3}\cdot\delta+ 2\cdot\delta^{2.5}}{9\sqrt{3}}}$. The inequality $\delta \ge \beta^1(3)$ is satisfied if and only if $\delta \in [D^3,1]$ and in this example, $D^3 \approx 0.361$. Hence, if $\delta \in [D^3,1]$, the regular strategy profile is a unique responsive PBE (ref. Theorem \ref{thm:char}). If $T=2$, then $\beta^1(2) = \sqrt{\frac{\delta}{3}}$ and $D^2 = \frac{1}{3}$. So, the regular strategy profile is a unique responsive PBE for $\delta \in [D^2,1]$. For $T \ge 2$, if $\delta \in (0, D^2)$, the game has no responsive PBE among the essentially Markov strategy profiles (ref. Theorem \ref{exist}). 

Now we consider the setting in which the game has infinite horizon. By Proposition \ref{sinc2}, the game has a unique regular strategy profile. In this profile, the receiver plays the obeying strategy and the sender plays the sincere strategy (against the receiver's strategy) with threshold $\beta$ at each period $t$. Here $\beta$ is the unique solution in $[0,1]$ to the equation $H(x) = x$. 

We have $H(\beta) = \beta$, which can be rewritten as $\delta = \frac{3\beta^2}{1+2\beta^3}$. For $\delta<1$, we have $\beta<1$. Recall from Section \ref{sec:infin} that the bound $D$ is defined as the smallest number such that $\delta \cdot V(1) \ge V(\beta)$ for $\delta \in [D,1]$. The inequality is equivalent to $\delta \ge \beta$. This holds if and only if $\delta \in [D,1]$. In this example, $D$ solves the equation $2\delta  x^3-3x^2+\delta=0$. Hence, $D \approx 0.366$. So, if $\delta \in [D,1)$, the regular strategy profile is a unique responsive PBE (ref. Theorem \ref{stat-thm}). 

If the payoffs are period independent, i.e. $\delta=1$, then by Theorem \ref{nonex}, there is no responsive PBE among the essentially Markov strategy profiles. Indeed, in this case, we have $\beta = 1$. So, according to the regular strategy profile, the sender sends the message $m_c$ whenever the state is less than 1 and the receiver obeys. Hence, the game is played forever with probability 1, and the expected payoff is 0 for both the players. 
\hfill \qed 
\end{example}

\begin{example} \rm
Here we assume $f(x) = x^2$ and $g(x) = x^3$.
As the payoff function of the sender is same as in Example 1, so is the unique regular strategy profile. As $g(x) = x^3$, we have $V(x) = \frac{x^3}{4}$. In this case, we have $D^T = 0$ for each $T \ge 1$ and $D = 0$. Hence, in the  setting of the game with a finite horizon $T$ (for $T \ge 1$), the regular strategy profile is the unique responsive PBE for $\delta \in (0,1]$ and in the  setting of the game with the infinite horizon, the regular strategy profile is the unique responsive PBE for $\delta \in (0,1)$.  
\hfill \qed 
\end{example}

\section{Concluding remarks}\label{sec:conclude}

This paper shows that the model of sender-receiver stopping games differs from the other models of dynamic sender-receiver games in the literature. The striking feature about this model stated by our main results is that under the responsive PBE the sender plays the threshold strategy optimal for himself and the receiver simply obeys. This is surprising, as the receiver has to comply with the sender regardless his own payoff function. Under the responsive PBE the sender gets the maximum possible payoff for himself. Hence, the delegation of the decision making to the receiver does not hurt the sender.

We see many interesting open question to be addressed in future work. One immediate question would be the extension to arbitrary distributions of the state of the world. The simple case in which nature draws an iid state from a strictly increasing continuous distribution is discussed in Appendix C. A challenging future extension would be the case in which the state of the world follows a Markov chain. We are currently investigating the situation with multiple senders, so that the receiver can make better informed decisions.

\section{The proofs}\label{proof}

\subsection{The proof of Theorem \ref{thm:char}}
We will prove Theorem \ref{thm:char} in two parts: Claim \ref{char1} and Claim \ref{char2}. So we fix a sender-receiver stopping game with finite horizon $T$, and with payoffs that are period-independent $(\delta=1)$ or they are discounted with discount factor $\delta\in(0,1)$.

\begin{claim} \label{char1}
Let $\delta \in [D^T,1]$. Suppose that $\tau$ is responsive.
If the strategy profile $(\sigma, \tau)$ is a PBE, then $(\sigma, \tau)$ is regular. 
\end{claim}

Proof. \quad
Fix $\delta \in [D^T,1]$. Suppose that $(\sigma, \tau)$ is a PBE with $\tau$ being responsive. We will prove that $(\sigma, \tau)$ is regular. We shall do so by proving that the sender's strategy $\sigma$ is sincere and the receiver's strategy $\tau$ is obedient at each period $t=1,\ldots,T-1$. We apply backward induction by considering the periods in the order $T-1,T-2,\ldots, 1$.  

We recall from Section \ref{str+exp} that at period $t$, the expected payoffs $U_s^{t}(\sigma, \tau)$ and  $U_r^{t}(\sigma, \tau)$ are functions of the histories of the sender and the receiver respectively.
 
For $t = T-1, T-2, \ldots, 1$, let $Q(t)$ be the following list of statements:
\begin{itemize}
\item[{[1]}] $\sigma$ is sincere at period $t$, and the corresponding threshold is equal to $\beta^t$. 
\item[{[2]}] $U_r^{t+1}(\sigma, \tau) > \frac{1}{\beta^t} \int_0^{\beta^t} g^t(\theta) d\theta$.
\item[{[3]}] $U_r^{t+1}(\sigma, \tau) < \frac{1}{1 - \beta^t} \cdot \int_{\beta^t}^1 g^t(\theta) d\theta$.
\item[{[4]}] $\tau$ is obedient at period $t$. That is, for each history $h_r^t$,
$\tau^{t}(h_r^t, m_c) = 1$ and $\tau^{t}(h_r^t, m_q) = 0$.
\item[{[5]}] $U^t_s(\sigma, \tau)$ and $U^t_r(\sigma, \tau)$ do not depend on history up to period $t$.
\end{itemize}
\vskip6pt

For ease of notation, and suppressing possible history dependence, for every period $t=1,\ldots,T$ let $p^t = \tau^t(h_r^t, m_c)$ be the probability on the action $a_c$ on seeing the message $m_c$,
and let $q^t = \tau^t(h_r^t, m_q)$ be the probability on the action $a_c$ on seeing the message $m_q$.
As $\tau$ is responsive, we know that $p^t > q^t$.
\vskip6pt

{\bf Remark on period $T$.} The final period is a special case. At this period the sender always sends message $m_q$ and the receiver always chooses action $a_q$. So, $U_s^{T}(\sigma,\tau)$ and $U_r^{T}(\sigma,\tau)$ are independent of the history up to period $T$.
\vskip6pt

\textbf{The induction step.} We assume that $Q(t+1),\ldots,Q(T-1)$ are true, for some $t\in\{1,\ldots,T-1\}$. We show that $Q(t)$ is also true.

\textbf{Item [1] of Q(t).} First we show that $\beta^t$ is the (unique) solution to the equation $f^t(x) = U_s^{t+1}(\sigma,\tau)$. For $t=T-1$ we have 
\begin{eqnarray*}
U_s^{T}(\sigma,\tau) & = & \int_0^1 f^{T}(\theta) d\theta \cr
& = & \delta^{T-1}\cdot \int_0^1 f(\theta) d\theta = \delta^{T-2}\cdot f(H(0)) \cr
& = & \delta^{T-2}\cdot f(H(\beta^T)) = \delta^{T-2}\cdot f(\beta^{T-1}) \cr
& = & f^{T-1}(\beta^{T-1}).
\end{eqnarray*}


Now consider the case when $t\leq T-2$. By item [5] of $Q(t+1)$, the expected payoff $U_s^{t+1}(\sigma,\tau)$ is independent of the history up to period $t+1$. By item [5] of $Q(t+2)$ when $t\leq T-3$ and by the remark on period $T$ when $t=T-2$, the expected payoff $U_s^{t+2}(\sigma,\tau)$ is independent of the history up to period $t+2$. We have
\begin{eqnarray*}
U_s^{t+1}(\sigma,\tau) & = &
\beta^{t+1} \cdot U_s^{t+2}(\sigma,\tau) + \int_{\beta^{t+1}}^1 f^{t+1}(\theta) d\theta \cr
& = &
\beta^{t+1} \cdot f^{t+1}(\beta^{t+1})  + \int_{\beta^{t+1}}^1 f^{t+1}(\theta) d\theta \cr
& = &
\delta^{t-1} \cdot \delta \cdot \left[ \beta^{t+1} \cdot f(\beta^{t+1})  + \int_{\beta^{t+1}}^1 f(\theta) d\theta \right] \cr
& = & \delta^{t-1}\cdot f(H(\beta^{t+1})) \cr
& = & \delta^{t-1}\cdot f(\beta^t) \cr
& = & f^t(\beta^t).
\end{eqnarray*}
For the first equality, we use items [1] and [4] of $Q(t+1)$. For the second equality, we use item [1] of $Q(t+1)$ and the definition of sincere strategy. For the fifth equality, we use the definition of $\beta^t$. 
\vskip6pt

Thus, when $\theta^t < \beta^t$, the sender expects to get a strictly better payoff if the receiver continues than when the receiver quits.
As $p^t > q^t$, the sender strictly prefers to send the message $m_c$ over $m_q$ if $\theta^t < \beta^t$.
By the same reasoning, the sender strictly prefers to send the message $m_q$ over $m_c$ if $\theta^t > \beta^t$.
Hence, $\sigma$ is sincere at period $t$ with threshold value $\beta^t$.
\vskip6pt

\textbf{Item [2] of Q(t).} 
Quitting at period $t+1$ guarantees payoff $\int_0^1 g^{t+1} (\theta) d\theta$ for the receiver.
So by using Lemma \ref{lem:fin}.2, we get
$$
\frac{1}{\beta^t} \cdot \int_0^{\beta^t} g^t(\theta) d\theta \, < \,
\int_0^1 g^{t+1} (\theta) d\theta \,\le\, U_r^{t+1}(\sigma, \tau).
$$
This proves item [2] of $Q(t)$.
\vskip6pt

\textbf{Item [3] of Q(t).} 
For $t=T-1$, by using Lemma \ref{lem:obs}, we have 
$$
U_r^T(\sigma, \tau) \,=\, \int_0^1 g^T(\theta) d\theta \,\leq\, \int_0^1 g^{T-1}(\theta) d\theta \,<\, \frac{1}{1 - \beta^{T-1}} \cdot \int_{\beta^{T-1}}^1 g^{T-1}(\theta) d \theta.
$$

For $t<T-1$ we have
\begin{eqnarray*}
U_r^{t+1}(\sigma, \tau)
& = &
\beta^{t+1} \cdot U_r^{t+2}(\sigma, \tau) + \int_{\beta^{t+1}}^1 g^{t+1}(\theta) d\theta \cr
& < &
\frac{1}{1 - \beta^{t+1}} \cdot \int_{\beta^{t+1}}^1 g^{t+1}(\theta) d\theta \cr 
& \le &
\frac{1}{1 - \beta^{t+1}} \cdot \int_{\beta^{t+1}}^1 g^t(\theta) d\theta \cr 
& < &
\frac{1}{1 - \beta^t} \cdot \int_{\beta^t}^1 g^t(\theta) d \theta.
\end{eqnarray*}
In the equality, we use items [1] and [4] of $Q(t+1)$. In the first inequality, we use item [3] of $Q(t+1)$. For the last inequality, we use Lemmas \ref{lem:fin}.1 and \ref{lem:obs}.
\vskip6pt

\textbf{Item [4] of Q(t).} 
Assume the message is $m_c$ at period $t$. Then by item [1] of $Q(t)$, we have $\theta^t \in [0,\beta^t]$. We want to show that $\tau$ plays action $a_c$ at period $t$. 
The expected payoff for the receiver on quitting is $\frac{1}{\beta^t} \int_0^{\beta^t} g^t(\theta) d\theta$. By item [2] of $Q(t)$ we have 
$$
U_r^{t+1}(\sigma, \tau) > \frac{1}{\beta^t} \int_0^{\beta^t} g^t(\theta) d\theta.
$$
Since $(\sigma,\tau)$ is a PBE, $\tau$ has to play action $a_c$ at period $t$. 
\vskip6pt

Now, assume the message is $m_q$ at period $t$. Then by item [1] of $Q(t)$, we have $\theta^t \in [\beta^t,1]$. We want to show that $\tau$ plays action $a_q$ at period $t$. 
The expected payoff for the receiver on quitting is $\frac{1}{1-\beta^t} \int_{\beta^t}^1 g^t(\theta) d\theta$. By [3] of $Q(t)$ we have 
$$
U_r^{t+1}(\sigma, \tau) < \frac{1}{1-\beta^t} \int_{\beta^t}^1 g^t(\theta) d\theta.
$$
Since $(\sigma,\tau)$ is a PBE, $\tau$ has to play action $a_q$ at period $t$.
\vskip6pt

\textbf{Item [5] of Q(t).} 
By using items [1] and [4] of $Q(t),\ldots,Q(T-1)$ and the remark on period $T$, the statement follows immediately.
\hfill \qed \vskip6pt

\begin{claim} \label{char2}
Let $\delta \in [D^T,1]$. Then the regular strategy profile is a PBE. 
\end{claim}

Proof. \quad
Let $(\sigma, \tau)$ be regular. Then the expected payoffs $U_s^t(\sigma,\tau)$ and $U_r^t(\sigma,\tau)$ are independent of the history up to period $t-1$ for all periods $t=1,\ldots,T$. We show that $(\sigma, \tau)$ is a PBE in four steps. 
\vskip6pt

\textbf{Step 1.} \quad We show that, for all $t=1,\ldots,T$, the threshold used by $\sigma$ at period $t$ is exactly $\beta^t$. \vskip6pt

Proof of step 1. \quad
The proof is by backward induction. At period $T$, by definition, $\sigma$ uses threshold $\beta^T=0$.

Now assume that at some period $t=2,\ldots,T$, the strategy $\sigma$ uses threshold $\beta^t$. We argue that the threshold of $\sigma$ at period $t-1$ is $\beta^{t-1}$.  We have
\begin{eqnarray*}
U_s^{t}(\sigma, \tau)
& = &
\beta^{t} \cdot U_s^{t+1}(\sigma,\tau)+ \int_{\beta^{t}}^1 f^{t}(\theta) d\theta \cr
& = &
\beta^{t} \cdot f^t(\beta^{t})+ \int_{\beta^{t}}^1 f^{t}(\theta) d\theta \cr
& = &
\delta^{t-1} \cdot \Big[\beta^{t} \cdot f(\beta^{t})+ \int_{\beta^{t}}^1 f(\theta) d\theta \Big] \cr
& = &
\delta^{t-1} \cdot G(\beta^t) = \delta^{t-2} \cdot f(H(\beta^{t}))\cr
& = &
\delta^{t-2} \cdot f(\beta^{t-1}) = f^{t-1}(\beta^{t-1}). 
\end{eqnarray*}
In the first equality, we use that $\sigma$ uses threshold $\beta^t$ and $\tau$ is obedient at period $t$. In the second equality, we use that $\sigma$ is sincere at period $t$. In the sixth equality, we use the definition of $\beta^{t-1}$.  

Since $\sigma$ is sincere at period $t-1$, the threshold at period $t-1$ must be $\beta^{t-1}$, as desired. This completes the proof of step 1.
\vskip6pt

\textbf{Step 2.} \quad For each period $t=1,\ldots,T-1$, let $Q(t)$ be the statement that the following conditions hold:
\begin{enumerate}
	\item[{[1]}] If the receiver gets the message $m_q$ at period $t$, then 
$$
U_r^{t+1}(\sigma, \tau) < \frac{1}{1 - \beta^t} \cdot \int_{\beta^t}^1 g^t(\theta) d\theta.
$$
\item[{[2]}] If the receiver gets the message $m_c$ at period $t$, then 
$$
U_r^{t+1}(\sigma, \tau) > \frac{1}{\beta^t} \int_0^{\beta^t} g^t(\theta) d\theta.
$$
\end{enumerate}
We prove that $Q(t)$ holds for all $t=1,\ldots,T-1$.
\vskip6pt

Proof of step 2. \quad The proof is by backward induction. First we consider period $T-1$. We have  by Lemma \ref{lem:obs}
\[U_r^{T}(\sigma, \tau) =\int_0^1 g^T(\theta) d\theta  \le \int_0^1 g^{T-1}(\theta) d\theta < \frac{1}{1 - \beta^{T-1}} \cdot \int_{\beta^{T-1}}^1 g^{T-1}(\theta) d\theta, \]

which proves item [1] of the statement $Q(T-1)$. Item [2] of the statement $Q(T-1)$ follows from Lemma \ref{lem:fin}.2.

Now assume that $Q(t+1)$ is true, where $t \in\{1,\ldots,T-2\}$. We prove that $Q(t)$ also holds. 

First we prove the item [1] of $Q(t)$. We have
\begin{eqnarray*}
U_r^{t+1}(\sigma, \tau) & = & \beta^{t+1} \cdot U_r^{t+2}(\sigma, \tau) + \int_{\beta^{t+1}}^1 g^{t+1}(\theta) d\theta \cr
& \le &
\beta^{t+1} \cdot \frac{1}{1 - \beta^{t+1}} \cdot \int_{\beta^{t+1}}^1 g^{t+1}(\theta) d\theta + \int_{\beta^{t+1}}^1 g^{t+1}(\theta) d\theta \cr 
& = &
\frac{1}{1 - \beta^{t+1}} \cdot \int_{\beta^{t+1}}^1 g^{t+1}(\theta) d\theta \cr 
& \le &
\frac{1}{1 - \beta^{t+1}} \cdot \int_{\beta^{t+1}}^1 g^{t}(\theta) d\theta \cr 
& < &
\frac{1}{1 - \beta^{t}} \cdot \int_{\beta^{t}}^1 g^{t}(\theta) d \theta.
\end{eqnarray*}
In the first equality, we use that, by step 1, $\sigma$ uses threshold $\beta^{t+1}$ and $\tau$ is obedient at period $t+1$. In the first inequality, we use item [1] of $Q(t+1)$. In the last inequality, we use Lemma \ref{lem:obs} and the fact that $\beta^t > \beta^{t+1}$. This completes the proof of item [1] of $Q(t)$.

Now we prove the item [2] of $Q(t)$. We have
\begin{eqnarray*}
U_r^{t+1}(\sigma, \tau)
& = &\beta^{t+1} \cdot U_r^{t+2}(\sigma, \tau) + \int_{\beta^{t+1}}^1 g^{t+1} (\theta) d\theta  \cr 
& \ge &\beta^{t+1} \cdot \frac{1}{\beta^{t+1}} \cdot \int_0^{\beta^{t+1}} g^{t+1}(\theta) d\theta + \int_{\beta^{t+1}}^1 g^{t+1} (\theta) d\theta  \cr 
& = &
\int_0^1 g^{t+1} (\theta) d\theta \cr
& > &
\frac{1}{\beta^t} \cdot \int_0^{\beta^t} g^t(\theta) d\theta.
\end{eqnarray*}
In the first equality, we use that, by step 1, $\sigma$ uses threshold $\beta^{t+1}$ and $\tau$ is obedient at period $t+1$.  In the first inequality, we use item [2] of $Q(t+1)$. In the second inequality, we use the Lemma \ref{lem:fin}.2. This completes the proof of item [2] of $Q(t)$.

\textbf{Step 3.} We prove that, starting from any period $t=1,\ldots,T$, the obedient strategy $\tau$ of the receiver is a best response against the sincere strategy $\sigma$ of the sender. 
\vskip6pt

Proof of step 3. At any period $t \in \{1,\ldots,T-1\}$, consider the case where the receives gets message $m_q$. We show that it is a (unique) best response for the receiver to play $a_q$. That is, $\tau$ plays a best response at period $t$.

Because the receiver gets the message $m_q$ at period $t$, we have by step 1 that $\theta^t \in [\beta^t,1]$.
So, the expected payoff for the receiver on quitting is $\frac{1}{1-\beta^t} \int_{\beta^t}^1 g^t(\theta) d\theta$.
By step 2, we have 
$$
U_r^{t+1}(\sigma, \tau) < \frac{1}{1 - \beta^t} \cdot \int_{\beta^t}^1 g^t(\theta) d\theta.
$$
Hence, it is a best response for the receiver to play $a_q$.

Now at any period $t \in \{1,\ldots,T-1\}$, consider the case where the receives gets message $m_c$. We show that it is a (unique) best response for the receiver to play $a_c$. That is, $\tau$ plays a best response at period $t$.

Because the receiver gets the message $m_c$ at period $t$, we have by step 1 that $\theta^t \in [0,\beta^t]$. So, the expected payoff for the receiver on quitting is $\frac{1}{\beta^t} \int_0^{\beta^t} g^t(\theta) d\theta$.
By step 2, we have 
$$
U_r^{t+1}(\sigma, \tau) > \frac{1}{\beta^t} \cdot \int_0^{\beta^t} g^t(\theta) d\theta.
$$
Hence, it is a best response for the receiver to play $a_c$.\vskip6pt

\textbf{Step 4.} As the final step, we argue that $(\sigma,\tau)$ is a PBE. 

Proof of step 4. By step 3, at any period, $\tau$ is best response against $\sigma$.
Conversely, we show that $\sigma$ is a best response against $\tau$.
Recall that for each period $t=1,\ldots,T-1$, the expected payoff $U_s^{t+1}(\sigma,\tau)$ is history independent. 
Since $\tau$ is obedient, the sender receives payoff $f^t(\theta^t)$ if $\sigma$ sends $m_q$ at period $t$ and payoff $U_s^{t+1}(\sigma,\tau)$ if $\sigma$ sends $m_c$.
So, the sender plays a best response at period $t$ when he sends $m_c$ if $f^t(\theta^t) < U_s^{t+1}(\sigma,\tau)$
and $m_q$ if $f^t(\theta^t) > U_s^{t+1}(\sigma,\tau)$. Hence, since $U_s^{t+1}(\sigma,\tau)=f^t(\beta^t)$, the sender plays a best response at period $t$ when he sends $m_c$ if $\theta^t < \beta^t$ and $m_q$ if $\theta^t > \beta^t$. Thus, by step 1, the strategy $\sigma$ is a best response against $\tau$ at period $t$. This completes the proof of the theorem. \hfill \qed

\subsection{The proof of Theorem \ref{exist}}

In this section we prove Theorem \ref{exist}. So, let $f$ and $g$ be strictly increasing functions from $I$ to $\re_+$ with $f(0)=g(0)=0$. Consider a finite horizon $T\geq 2$. Assume by way of contradiction that a strategy profile $(\sigma, \tau)$ is a PBE where $\tau$ is a responsive strategy. 

As the receiver must quit at period $T$, we have $\:U_s^T(\sigma, \tau) = \int_0^1 \delta^{T-1} f(\theta) d\theta. \:$ Let $\:\alpha^{T-1} = f^{-1} \big(\delta \cdot \int_0^1  f(\theta) d\theta \big). \:$ At period $T-1$, if the receiver quits, the sender obtains the payoff $f^{T-1}(\theta^{T-1})$ and if the receiver continues, the sender obtains the expected payoff $U_s^T(\sigma, \tau)$. As $\tau$ is responsive, the sender strictly prefers to send the message $m_c$ if $\theta^{T-1} < \alpha^{T-1}$ and the message $m_q$ if $\theta^{T-1} > \alpha^{T-1}$. Hence, the strategy $\sigma$ has a threshold $\alpha^{T-1}$ at period $T-1$. 

Assume that the receiver gets the message $m_c$ at period $T-1$. As $\alpha^{T-1} > 0$, this happens with positive probability. If the receiver continues, he obtains the expected payoff 
$$ U_r^T(\sigma, \tau) = \int_0^1 \delta^{T-1} g(\theta) d\theta \: = \: \delta^{T-1} \cdot V(1).$$ And if the receiver quits, he obtains the expected payoff $$ \frac{1}{\alpha^{T-1}}\int_0^{\alpha^{T-1}} \delta^{T-2} g(\theta) d\theta \: = \: \delta^{T-2} \cdot V(\alpha^{T-1}). $$ 

As $\delta < D^2$, we have $\delta \cdot V(1) <  V(\alpha^{T-1})$. So, the receiver strictly prefers to quit on receiving the message $m_c$. This is however in contradiction with $\tau$ being responsive. So, the game admits no responsive PBE.
\hfill \qed

\subsection{The proof of Theorem \ref{stat-thm}}

We will prove Theorem \ref{stat-thm} in two parts: Claim \ref{char1-infinite} and Claim \ref{char2-infinite}. So we fix a sender-receiver stopping game with infinite horizon and with payoffs that are discounted with discount factor $\delta\in(0,1)$.

\begin{claim}\label{char1-infinite}
Suppose that $\delta  \in (D,1)$. Let $(\sigma, \tau)$ be an essentially Markov PBE where $\tau$ is responsive. Then, $(\sigma, \tau)$ is the regular strategy profile. 
\end{claim}

Proof. \quad
Suppose that $\delta  \in (D,1)$. Let $(\sigma, \tau)$ be an essentially Markov PBE where $\tau$ is responsive.

Again, let $p^t = \tau^t(h_r^t, m_c)$ be the probability on the action $a_c$ on seeing the message $m_c$,
and let $q^t = \tau^t(h_r^t, m_q)$ be the probability on the action $a_c$ on seeing the message $m_q$. Since $\tau$ is responsive, $p^t > q^t$ for each $t$.
We prove the statement in a series of steps.
\vskip6pt

\textbf{Step 1.} \quad
The strategy $\sigma$ has threshold $\alpha^t$ at each period $t \in \mathbb{N}$ that satisfies $f^t(\alpha^t) = U^{t+1}_s(\sigma, \tau)$. Hence, $\sigma$ is sincere.  
\vskip6pt

Proof of step 1. \quad
Fix a period $t \in \mathbb{N}$. Since $(\sigma, \tau)$ is essentially Markov, $U_s^{t+1}(\sigma,\tau)$ is independent of the history up to period $t+1$. Further, note that $f^t(1) > f^{t+1}(1) \ge  U_s^{t+1}(\sigma,\tau) \ge 0=f^t(0)$. So, the equation $f^t(x) = U_s^{t+1}(\sigma,\tau)$ has the unique solution, say $\alpha^t\in[0,1)$.
\vskip6pt

Assume first that $\theta^t < \alpha^t$. Then, $f^t(\theta^t) < f^t(\alpha^t)=U_s^{t+1}(\sigma,\tau)$. Thus, the sender expects to get a strictly better payoff when the receiver continues than when the receiver quits. As $p^t > q^t$, the sender strictly prefers to send the message $m_c$ over $m_q$. Because $(\sigma,\tau)$ is a PBE, $\sigma$ sends the message $m_c$ at period $t$.

Assume now that $\theta^t > \alpha^t$. Then, $f^t(\theta^t) > f^t(\alpha^t)=U_s^{t+1}(\sigma,\tau)$. By the same reasoning, $\sigma$ sends the message $m_q$ at period $t$.
Thus, $\sigma$ is sincere at period $t$ with threshold $\alpha^t$. This completes the proof of step 1.
\vskip6pt

\textbf{Step 2.} \quad Define for each period $t\in\mathbb{N}$
$$
K(t) = \frac{1}{\alpha^t} \cdot \int_0^{\alpha^t} g^t(\theta) d\theta
\quad \hbox{and} \quad
L(t) = \frac{1}{1 - \alpha^t} \cdot \int_{\alpha^t}^1 g^t(\theta) d\theta
$$
Then, for every $t\in\mathbb{N}$ we have $K(t) \le U_r^{t+1}(\sigma, \tau) \le L(t)$.
\vskip6pt

Proof of step 2. \quad
Fix $t\in\mathbb{N}$. As $\tau$ is responsive, we know that $1 \ge p^t > q^t \ge 0$.
We know that at period $t$, the strategy $\sigma$ uses threshold $\alpha^t$. So, $K(t)$ is the expected payoff to the receiver when he quits, upon getting the message $m_c$. Since $p^t > 0$, it follows that $K(t)\le U_r^{t+1}(\sigma, \tau)$. Similarly, $L(t)$ is the expected payoff to the receiver when he quits, upon getting the message $m_q$. Since $q^t < 1$, it follows that $U_r^{t+1}(\sigma, \tau) \le L(t)$. This completes the proof of step 2.
\vskip6pt

\textbf{Step 3.} \quad
For each period $t\in\mathbb{N}$ we have $\alpha^t \le \beta$.
\vskip6pt

Proof of Step 3. \quad
Fix $t \in \mathbb{N}$. Suppose by way of contradiction that $\alpha^t > \beta$. Note that
\begin{eqnarray*}
f^t(\alpha^t)
& = &
U_s^{t+1}(\sigma,\tau) \cr
& \le &
\alpha^{t+1} \cdot U_s^{t+2}(\sigma,\tau) + \int_{\alpha^{t+1}}^1 f^{t+1}(\theta) d\theta \cr
& = &
\alpha^{t+1} \cdot f^{t+1}(\alpha^{t+1})  + \int_{\alpha^{t+1}}^1 f^{t+1}(\theta) d\theta \cr
& = &
\delta^t \cdot G(\alpha^{t+1}).
\end{eqnarray*}
The first and second equalities follow from the definitions of $\alpha^{t}$ and $\alpha^{t+1}$ and the last equality from the definition of $G$. The inequality can be explained as follows. If $\theta^{t+1} < \alpha^{t+1}$, which has probability $\alpha^{t+1}$, then the sender's payoff is at most $U_s^{t+2}(\sigma,\tau)$ (which is realized if the receiver continues). If $\theta^{t+1} > \alpha^{t+1}$, which has probability $1-\alpha^{t+1}$, then the sender's payoff is at most $\frac{1}{1-\alpha^{t+1}} \int_{\alpha^{t+1}}^1 f^{t+1}(\theta) d\theta$ (which is realized if the receiver quits). 
\vskip6pt

It follows that $f(\alpha^t) \le \delta \cdot G(\alpha^{t+1})$. So, $\,\alpha^t \le H(\alpha^{t+1}).\,$ As $\,\alpha^t > \beta\,$ by assumption, by Lemma \ref{lem:func}.3, $\,H\big(\alpha^t\big) < \alpha^t \le H(\alpha^{t+1}).\,$  This implies $\,\alpha^t < \alpha^{t+1}\,$ and so $\,\alpha^{t+1} > \alpha.\,$ Repeating the whole argument for $t+1$, we get $\,\alpha^{t+2} > \alpha^{t+1} > \beta.\,$ And hence iterating the argument, we can conclude that the sequence $(\alpha^{t'})_{t'=t}^{\infty}$ is strictly increasing. By definition, $\,\alpha^t \le 1\,$ for each $t \in \mathbb{N}$. Moreover, $\,\alpha^{t'} \le H\big(\alpha^{t'+1}\big) \le H(1)$ for each $t'>t$. Hence, the sequence $(\alpha^{t'})_{t'=t}^{\infty}$ converges to some $r \le H(1) < 1$.
\vskip6pt

Denote $\,z =$ min $\{x-H(x) : x \in [\alpha^t,1]\}.\,$ By Lemma \ref{lem:func}.3, $\,H(x) < x\,$ for $\,x \ge \alpha^t > \beta.\,$ By continuity of $H$, we have $z>0$. As the sequence $(\alpha^{t'})_{t'=t}^\infty$ converges to $r$, for $\epsilon = \frac{z}{2}$, there exists $N>0$ such that $\alpha^{t'} \in [r-\epsilon,r]$ for all $t'>N$. Now for any fixed $t'>N$, we have  $\,\alpha^{t'+1} \, \ge  \,  H\big(\alpha^{t'+1}\big) + z \, \ge \, \alpha^{t'} + z \, \ge \, r- \epsilon + z \, = \, r+ \frac{z}{2}$, hence $\alpha^{t'+1} \ge r$. This is a contradiction. This completes the proof of step 3.
\vskip6pt

\textbf{Step 4.} \quad
The strategy $\tau$ is obedient.
\vskip6pt

Proof of Step 4. \quad\\
Step 4.1: \quad
We show that $\tau^t(h_r^t,m_c) = p^t = 1$ for every period $t$ and history $h_r^t$ of the receiver. Take any $t\in\mathbb{N}$. It is sufficient to show that  $K(t) < U_r^{t+1}(\sigma, \tau)$.
 By step 3, $\alpha^t \le \beta$. It holds that
 \[K(t) \,=\, \frac{1}{\alpha^t} \cdot \int_0^{\alpha^t} g^t(\theta) d\theta \,\le\,\frac{1}{\beta} \cdot \int_0^{\beta} g^t(\theta) d\theta \,< \,\int_0^1 g^{t+1} (\theta) d\theta \,\le\,
U_r^{t+1}(\sigma, \tau).\]
In the first inequality, we use Lemma \ref{lem:obs}. In the second inequality, we use Lemma \ref{lem:infin}.3 and the assumption $\delta \cdot V(1) > V(\beta)$. Due to $(\sigma,\tau)$ being a PBE, the last inequality follows because quitting at period $t+1$ cannot be better for the receiver than playing $\tau$ against $\sigma$. This completes the proof of step 4.1.
\vskip6pt

Step 4.2: \quad
We show that $\tau^t(h_r^t,m_q) = q^t = 0$ for every period $t$ and history $h_r^t$ of the receiver. For every $t\in\mathbb{N}$, let $Q(t)$ be the the statement $U_r^{t+1}(\sigma, \tau) < L(t)$. As $(\sigma,\tau)$ is a PBE, $Q(t)$ implies that $q^t = 0$. It is sufficient to show that $Q(t)$ is true for every $t\in\mathbb{N}$.
\vskip6pt

Step 4.2.1: \quad
We show that $Q(t+1)$ implies $Q(t)$. Suppose that $Q(t+1)$ is true, so $q^{t+1} = 0$. Together with step 4.1, it follows that $\tau$ is obedient at period $t+1$. Then
\begin{eqnarray*}
f^t(\alpha^t) &=& U_s^{t+1}(\sigma,\tau) \cr
& = &
\alpha^{t+1} \cdot U_s^{t+2}(\sigma,\tau) + \int_{\alpha^{t+1}}^1 f^{t+1}(\theta) d\theta \cr
& = &
\alpha^{t+1} \cdot f^{t+1}(\alpha^{t+1})  + \int_{\alpha^{t+1}}^1 f^{t+1}(\theta) d\theta \cr
& = &
\delta^t \cdot G(\alpha^{t+1}).
\end{eqnarray*}
The first and third equalities follow from the definitions of $\alpha^{t}$ and $\alpha^{t+1}$ and the last equality from the definition of $G$. The second equality follows from $\tau$ being obedient at period $t+1$.

It follows that $\alpha^t = H(\alpha^{t+1})$. By step 3, we have $\alpha^t \le \beta$. Hence by Lemma \ref{lem:func}.3, $\alpha^t \ge \alpha^{t+1}$. Then,
\begin{eqnarray*}
U_r^{t+1}(\sigma, \tau)
& = &
\alpha^{t+1} \cdot U_r^{t+2}(\sigma, \tau) + \int_{\alpha^{t+1}}^1 g^{t+1}(\theta) d\theta \cr
& < &
\frac{\alpha^{t+1}}{1 - \alpha^{t+1}} \cdot \int_{\alpha^{t+1}}^1 g^{t+1}(\theta) d\theta + \int_{\alpha^{t+1}}^1 g^{t+1}(\theta) d\theta \cr
& = &
\frac{1}{1 - \alpha^{t+1}} \cdot \int_{\alpha^{t+1}}^1 g^{t+1}(\theta) d\theta \cr
& \le &
\frac{1}{1 - \alpha^t} \cdot \int_{\alpha^t}^1 g^{t+1}(\theta) d\theta \cr
& < &
\frac{1}{1 - \alpha^t} \cdot \int_{\alpha^t}^1 g^t(\theta) d\theta \cr
&=& L(t).
\end{eqnarray*}
In the first inequality we use that the statement $Q(t+1)$ is true. In the second inequality we use Lemma \ref{lem:obs}, which is applicable as $\alpha^t \ge \alpha^{t+1}$ by step 3. This completes the proof of step 4.2.1.
\vskip6pt

Step 4.2.2 \quad
We prove that $Q(t)$ is true for every $t$. Assume by way of contradiction that there is a $t \in \mathbb{N}$ for which $Q(t)$ is not true. By step 4.2.1, $Q(t')$ is not true for all $t' \ge t$. Then, by step 2, $U_r^{t'+1}(\sigma, \tau) = L(t')$ for all $t' \ge t$. Denote by $E^{t+1}_r$ the expected payoff of the receiver conditional on getting the message $m_q$ at period $t+1$. We have 
\begin{eqnarray*}
E^{t+1}_r
& = &
q^{t+1} \cdot U^{t+2}_r(\sigma, \tau) + (1 - q^{t+1}) \cdot L(t+1) \cr
& = &
q^{t+1} \cdot U^{t+2}_r(\sigma, \tau) + (1 - q^{t+1}) \cdot U^{t+2}_r(\sigma, \tau) \cr
& = &
U^{t+2}_r(\sigma, \tau).
\end{eqnarray*}

The first equality can be explained as follows. On receiving the message $m_q$ at period $t+1$, the strategy $\tau$ continues with probability $q^{t+1}$ and on continuing the receiver gets the payoff $U_r^{t+2}(\sigma, \tau)$, whereas $\tau$ quits with probability $1-q^{t+1}$ and on quitting the receiver gets the payoff $L(t+1)$. It follows that
$$
U_r^{t+1}(\sigma, \tau) = \alpha^{t+1} \cdot U_r^{t+2}(\sigma, \tau) + (1 - \alpha^{t+1}) \cdot E^{t+1}_r = U^{t+2}_r(\sigma, \tau).
$$

In the first equality we use the step 4.1. 

So, $U^{t+1}_r(\sigma, \tau) = U^{t+2}_r(\sigma, \tau)$. Iterating this argument implies that $U^{t+1}_r(\sigma, \tau) = U^{t+j}_r(\sigma, \tau)$ for any $j$. As $U_r^{t+j} \le g^{t+j}(1) = \delta^{t+j-1}g(1)$,
it then follows that $U^{t+1}_r(\sigma, \tau) = 0$. However, if the receiver quits at period $t+1$ instead, regardless of the message, then he receives the expected payoff $\int_0^1 g^{t+1}(\theta) d\theta > 0$. This contradicts the assumption that $(\sigma,\tau)$ is PBE. This completes the proof of step 4.2.2.
\vskip6pt

Now Claim \ref{char1-infinite} follows immediately from Proposition \ref{sinc2}.
\hfill \qed

\begin{claim}\label{char2-infinite}
Suppose that $\delta \in (D,1)$. Then the regular strategy profile is a PBE. 
\end{claim}

Proof. \quad Let $\delta \in (D,1)$ and let $(\sigma, \tau)$ be the regular strategy profile. From Proposition \ref{sinc2}, $\sigma$ is stationary with threshold $\beta$. Clearly, $U^{t+1}_r(\sigma, \tau) = \delta \cdot U^t_r(\sigma, \tau)$ for all periods $t\in\mathbb{N}$.
We show that $(\sigma, \tau)$ is a PBE. 
\vskip6pt

Since $(\sigma,\tau)$ is regular, we have 
$$
U_r^t(\sigma, \tau) = \beta \cdot U_r^{t+1}(\sigma, \tau) + \int_{\beta}^1 g^t(\theta) d\theta=\beta \cdot \delta\cdot U_r^{t}(\sigma, \tau) + \int_{\beta}^1 g^t(\theta) d\theta.
$$
It follows that
\begin{equation}\label{closed-form}
U_r^t(\sigma,\tau) = \frac{\delta^{t-1}}{1 - \delta \cdot \beta} \cdot \int_{\beta}^1 g(\theta) d \theta.
\end{equation}
Now we show that the receiver prefers to play $a_q$ on seeing $m_q$ at period $t$. For this, we need to show that
$$
U_r^{t+1}(\sigma, \tau) \le \frac{\delta^{t-1}}{1 - \beta} \int_{\beta}^1 g(\theta) d \theta.
$$
This follows easily from (\ref{closed-form}).
\vskip6pt

Finally, we show that the receiver prefers to play $a_c$ on seeing $m_c$ at period $t$. For this, we need to show that
$$
U_r^{t+1}(\sigma, \tau) \ge \frac{\delta^{t-1}}{\beta} \int_0^{\beta} g(\theta) d \theta.
$$
Using (\ref{closed-form}), the above inequality can be rewritten to
$$
\frac{1}{\beta} \cdot \int_0^{\beta} g(\theta) d \theta < \delta \cdot \int_0^1 g(\theta) d \theta. 
$$
This follows from Lemma \ref{lem:infin}.3 and the condition $V(\beta) < \delta \cdot V(1)$, which is due to $\delta\in(D,1)$.
\hfill \qed
\vskip6pt

\subsection{The proof of Theorem \ref{nonex}}
In this section, we prove Theorem \ref{nonex}. By way of contradiction, assume that $(\sigma,\tau)$ is an essentially Markov strategy profile and a responsive PBE. For every period $t$, let $p^t = \tau^t(h_r^t, m_c)$ be the probability on the action $a_c$ on seeing the message $m_c$,
and let $q^t = \tau^t(h_r^t, m_q)$ be the probability on the action $a_c$ on seeing the message $m_q$.
As $\tau$ is responsive, we know that $p^t > q^t$.
\vskip6pt

\textbf{Step 1.}
We prove that the sender's strategy $\sigma$ is a threshold strategy with some threshold $\alpha^t \in (0,1)$ at each period $t$.
\vskip6pt

Proof of step 1. \quad
Fix a period $t \in \mathbb{N}$. Since $(\sigma, \tau)$ is essentially Markov, $U_s^{t+1}(\sigma,\tau)$ is independent of the history up to period $t+1$. The equation $f(x) = U_s^{t+1}(\sigma,\tau)$ has a unique solution $\alpha^t\in[0,1]$. Notice that $f(0) <U_s^{t+1}(\sigma,\tau)<f(1)$ because state 0 are 1 have probability zero. Hence, $\alpha^t\in(0,1)$.
\vskip6pt

Assume first that $\theta^t < \alpha^t$. Then, $f(\theta^t) < f(\alpha^t)=U_s^{t+1}(\sigma,\tau)$. Thus, the sender expects to get a strictly better payoff when the receiver continues than when the receiver quits. As $p^t > q^t$, the sender strictly prefers to send the message $m_c$ over $m_q$. Because $(\sigma,\tau)$ is a PBE, $\sigma$ sends the message $m_c$ at period $t$.

Assume now that $\theta^t > \alpha^t$. Then, $f(\theta^t) > f(\alpha^t)=U_s^{t+1}(\sigma,\tau)$. By the same reasoning, $\sigma$ sends the message $m_q$ at period $t$.
This completes the proof of step 1.
\vskip6pt

\textbf{Step 2.}
We show that for every period $t$, we have $p^t = 1$.
\vskip6pt

Suppose that $\sigma$ sends the message $m_c$ at period $t$. Then, $\theta^t \le \alpha^t$.
If the receiver quits at period $t$, his expected payoff is $\int_0^{\alpha^t} g(\theta) d\theta$.
However, if the receiver continues at period $t$ and quits at period $t+1$ irrespective of the message by the sender,
his expected  payoff is $\int_0^1 g(\theta) d\theta$, which is strictly better because $\alpha^t < 1$ due to step 1.
So, action $a_q$ is not a best response for the receiver. It follows that $p^t = 1$.
\vskip6pt

\textbf{Step 3.}
We show that $\sum_{t=1}^\infty (1-q^t)=\infty$.
\vskip6pt

Let $z^t$ denote the probability under $(\sigma, \tau)$ that the receiver continues at period $t$, conditional on reaching period $t$. Thus, $z^t = \alpha^t \cdot p^t + (1-\alpha^t) \cdot q^t$.
\vskip6pt

We show that $\Pi_{t=1}^\infty z^t = 0$. Assume by way of contradiction that $\Pi_{t=1}^\infty z^t > 0$.
Because
$$
\Pi_{t=1}^\infty z^t = \Pi_{t=1}^n z^t \cdot \Pi_{t=n+1}^\infty z^t
$$
and $\Pi_{t=1}^n z^t \rightarrow \Pi_{t=1}^\infty z^t$ as $n \rightarrow \infty$, we have  $\Pi_{t=n+1}^\infty z^t \rightarrow 1$ as $n \to \infty$.
So, there is large period $t$ such that conditional on reaching period $t$, the receiver's expected payoff $U_r^t(\sigma, \tau)$ is less than $\int_0^1 g(\theta) d\theta$.
As the receiver can always guarantee an expected payoff of $\int_0^1 g(\theta) d\theta$ by simply quitting,
regardless the message sent by the sender, this is a contradiction. Hence, $\Pi_{t=1}^\infty z^t = 0$. 
\vskip6pt

Because $z^t = \alpha^t \cdot p^t + (1-\alpha^t) \cdot q^t$, by steps 1 and 2, we can conclude that $z^t > 0$.
Because $\Pi_{t=1}^\infty z^t = 0$, we obtain $\sum_{t=1}^\infty (1 - z^t) = \infty$. As $p^t > q^t$, we also have $z^t > q^t$.
Hence, $\sum_{t=1}^\infty (1-q^t) = \infty$.
\vskip6pt

\textbf{Step 4.} We derive a contradiction.
\vskip6pt

Let $\epsilon \in (0,1)$. We define a threshold strategy $\sigma_{\epsilon}$ for the sender as follows:
at each period $t$, if $\theta^t < 1-\epsilon$ then $\sigma_{\epsilon}$ sends the message $m_c$ and if
$\theta^t \ge 1- \epsilon$ then $\sigma_{\epsilon}$ sends the message $m_q$. 
\vskip6pt

We show that the sender's expected payoff under $(\sigma_{\epsilon}, \tau)$ is at least $f(1-\epsilon)$.
For this it is sufficient to prove that, under $(\sigma_{\epsilon}, \tau)$, with probability $1$ the receiver will eventually quit.
Let $z_{\epsilon}^t$ denote the probability under $(\sigma_{\epsilon}, \tau)$ that the receiver continues at period $t$,
conditional on reaching period $t$. By step 2, $z_{\epsilon}^t = (1 - \epsilon) \cdot p^t+ \epsilon \cdot q^t=(1 - \epsilon) + \epsilon \cdot q^t$.
Thus, by step 3, $\sum_{t=1}^\infty (1-z_{\epsilon}^t) = \epsilon \cdot \sum_{t=1}^\infty (1-q^t) = \infty$.
As $z_{\epsilon}^t > 0 $ for each $t$, we obtain $\Pi_{t=1}^\infty z_{\epsilon}^t = 0$.
Thus, under $(\sigma_{\epsilon}, \tau)$, with probability $1$ the receiver will eventually quit.
\vskip6pt

Since the sender can guarantee, for every $\epsilon \in (0,1)$, an expected payoff of at least $f(1-\epsilon)$ against $\tau$,
the sender's expected payoff under $(\sigma, \tau)$ must be at least $f(1)$. Since the state $1$ is realized with probability 0, this is a contradiction. This concludes the proof. \hfill \qed

\section*{Appendix}

\subsection*{A. The receiver's belief on the history of the sender}\label{condexp}

In this appendix, we describe the receiver's conditional probability distribution (or belief) $\mathbb{P}_{\sigma,\tau,h^t_r}$ on the set $H^t_s$ of possible histories for the sender, given the strategy profile $(\sigma,\tau)$ and the receiver's history $h_r^t = (m^1,m^2,\ldots, m^{t-1})$. 

Let $\sigma^k(m^k | h_s^k, \theta^k)$ denote the probability on the message $m^k$ under the strategy $\sigma$, given the history $h^k_s$ and the state $\theta^k$. For numbers $y^1,y^2,\ldots,y^{t-1} \in [0,1]$, the expression

$$ \displaystyle \int_0^{y^1} \int_0^{y^2} \ldots \int_0^{y^{t-1}} \:  \Bigg [ \prod_{k=1}^{t-1} \: \sigma^k(m^k | \theta^1,m^1,\ldots,\theta^{k-1},m^{k-1}, \theta^k) \Bigg ] \:\: d \theta^{t-1} \: d \theta^{t-2} \ldots \: d \theta^1. $$

is the probability of the event that $\theta^1 \le y^1$, $\theta^2\le y^2$,\ldots, $\theta^{t-1} \le y^{t-1}$ and the messages sent are $m^1,m^2,\ldots,m^{t-1}$. We denote this probability by $\chi^t_{(\sigma,\tau)} (h_r^t) (y^1,y^2,\ldots,y^{t-1})$.

The quantity $\chi^t_{(\sigma,\tau)} (h_r^t) (1,1,\ldots,1)$ is the probability that the history at period $t$ is $h_r^t$. Thus, the probability of the event that $\theta^1 \le y^1$, $\theta^2 \le y^2$, \ldots, $\theta^{t-1} \le y^{t-1}$ conditional on the messages $m^1,m^2,\ldots, m^{t-1}$ is
$$\Psi^t_{(\sigma,\tau)} (h_r^t)(y^1,y^2,\ldots,y^{t-1}) \, = \,  \frac{\chi^t_{(\sigma,\tau)} (h_r^t) (y^1,y^2,\ldots,y^{t-1})}{\chi^t_{(\sigma,\tau)} (h_r^t) (1,1,\ldots,1)}\,.$$

If a certain history $h_r^t$ occurs with probability zero, that is, if $\chi^t_{(\sigma,\tau)} (h_r^t) (1,1,\ldots,1) = 0$, then we define $\Psi^t_{(\sigma,\tau)}(h_r^t) $ to be any probability distribution. The choice of this probability distribution plays no role in our proofs. The probabilities $\Psi^t_{(\sigma,\tau)} (h_r^t)(y^1,y^2,\ldots,y^{t-1})$ induce the desired probability measure $\mathbb{P}_{\sigma,\tau,h^t_r}$ on the possible histories $h^t_s$ for the sender. 

\subsection*{B. Expected payoff} 

In this appendix, we provide the details of how the expected payoffs $U_s(\sigma,\tau)$, $U_r(\sigma,\tau)$ and the continuation expected payoffs $U_s^t(\sigma,\tau)(h_s^t)$ and $U_r^t(\sigma,\tau)(h_r^t)$ from period $t$ onward can be calculated.

It is both convenient and standard to assume that even if the receiver quits at some period $t$, play continues indefinitely,
but actions in any period beyond $t$ have no influence on the payoffs. With this assumption, a play of the game is a sequence $\omega = (\theta^t, m^t, a^t)_{t=1}^\infty$ where $\theta^t \in I$, $m^t\in M$ and $a^t \in A$. Denote by $\Omega = (I \times M \times A)^{\mathbb{N}} $ the set of all plays.
Given the usual Borel sigma-algebra of $I$, we endow $\Omega$ with the product sigma-algebra $\mathcal{B}$.

With abuse of notation, define $\theta^t \colon \Omega \to I$, $m^t \colon \Omega \to M$ and $a^t \colon \Omega \to A$ to be the projection maps from the set of plays, respectively to the state, the message and the action at period $t$. 
Let $S \colon \Omega \to \na \cup \{\infty\}$ be the mapping such that, for each $\omega \in \Omega$, $S(\omega)$ is the first period $t$ for which $a^t(\omega) = a_q$.
If there is no such $t$ then $S(\omega) = \infty$. It is the stopping time which indicates when the game effectively ends. For a play $\omega$, the payoffs for the players are given as follows
$$\Pi_s(\omega) = f^{S(\omega)}\Big(\theta^{S(\omega)}(\omega)\Big)\cdot \mathds{1}_{\{S(\omega) < \infty\}}, \hspace{4mm} \Pi_r(\omega) = g^{S(\omega)}\Big(\theta^{S(\omega)}(\omega)\Big)\cdot \mathds{1}_{\{S(\omega) < \infty\}}.$$

Any fixed strategy profile $(\sigma, \tau)$ induces a probability measure on the measurable space $(\Omega,\mathcal{B})$, denoted by $\mathbb{P}_{\sigma,\tau}$. 
The expectation with respect to this probability measure is denoted by $\mathbb{E}_{\sigma,\tau}$.
The expected payoff for the sender is given by $U_s(\sigma, \tau) = \mathbb{E}_{\sigma, \tau}\big[\Pi_s(\omega)\big]$
and the expected payoff for the receiver is given by $U_r(\sigma, \tau) = \mathbb{E}_{\sigma, \tau}\big[\Pi_r(\omega)\big]$.
\vskip6pt

Let $\Omega^{\ge t}$ denote the set of all continuation plays $\omega^{\ge t}=(\theta^k,m^k,a^k)_{k=t}^\infty$.
Given a history $h^t_s \in H^t_s$ for the sender, the continuation strategy $\sigma[h^t_s]=(\sigma^k[h^t_s])_{k=1}^\infty$ of $\sigma$ is defined in the usual way: for each period $k\in\mathbb{N}$, history $\overline{h}^k_s\in H^k_s$ and state $\theta^k\in I$ we let
\[\sigma^{k}[h^t_s](h^k_s,\theta^k) = \sigma^{t+k-1}(h_s^t,h^k_s,\theta^k).\]
Given a history $h^t_r \in H^t_r$ for the receiver, we define in a similar way the continuation strategy $\tau[h^t_r]=(\tau^k[h^t_r])_{k=1}^\infty$ of $\tau$. 

For each period $t$, let $\pi^t \colon H_s^t \to H_r^t$ be the map that projects the sender's history to the receiver's history. For a given history $h_s^t$ of the sender, the continuation strategies $\sigma[h_s^t]$ and $\tau[\pi(h_s^t)]$ induce a probability measure on the space $(\Omega,\mathcal{B})$, denoted by $\mathbb{P}_{\sigma,\tau,h_s^t}$. 
The expected continuation payoff for the sender is given by $U_s^t(\sigma, \tau)(h_s^t) = \mathbb{E}_{\sigma, \tau,h_s^t}\big[\Pi_s(\omega)\big]$. 

As discussed in Appendix A, the receiver has a probability distribution (belief) $\mathbb{P}_{\sigma,\tau,h^t_r}$ on the set $H_s^t$, conditional on his history $h_r^t$. The expected continuation payoff for the receiver is can be calculated as follows
\[U_r^t(\sigma,\tau)(h_r^t)\,=\,\int_{H^t_s} U_r(\sigma[h^t_s],\tau[h^t_r])\  \mathbb{P}_{\sigma,\tau,h^t_r}(dh^t_s).\]
Here, the integrand $U_r(\sigma[h^t_s],\tau[h^t_r])$ is the receiver's expected payoff given the continuation strategies $\sigma[h^t_s]$ and $\tau[h^t_r]$.
\vskip6pt

\subsection*{C.  Extension: Arbitrary distribution}

We consider an extension in which the states at each period are drawn from an arbitrary distribution for the games with finite or infinite horizon and with payoffs that are discounted or period independent. 

Consider a sender-receiver game where the payoffs are either discounted ($\delta<1$) or period independent ($\delta=1$). Let the characteristic functions $f$ and $g$ from $I$ to $\re_+$ be strictly increasing with $f(0)=g(0)=0$. At each period $t$, the state $\theta^t$ is drawn from a fixed cumulative distribution $F$ on $[0,1]$, independently from realized states of previous periods. We assume that $F$ is strictly increasing and continuous on $[0,1]$ and $F(0)=0$. We denote this game by $\mathcal{G}^F$. 

Using the game $\mathcal{G}^F$, we define a new game $\mathcal{G}^u$ with the same horizon $T$ in which the states at each period $t$ in the game $\mathcal{G}^u$ are drawn from the uniform distribution independently from states of previous periods. The game $\mathcal{G}^u$ has the same $\delta$ as the game $\mathcal{G}^F$ and has the characteristic functions $\hat{f}$ and $\hat{g}$ which are defined as follows: $\hat{f}(x) = f(F^{-1}(x))$, $\hat{g}(x) = g(F^{-1}(x))$. 

Given a strategy profile $(\sigma,\tau)$ in the game $\mathcal{G}^F$, consider a strategy profile $(\hat{\sigma},\hat{\tau})$ in the game $\mathcal{G}^u$, defined as follows: $\hat{\sigma}^t(\theta^t) = \sigma^t(F^{-1}(\theta^t))$ and $\hat{\tau}^t(m^t) = \tau^t(m^t)$. It is straightforward, but tedious to show that the payoffs of the players in the game $\mathcal{G}^u$ when the strategy profile is $(\hat{\sigma},\hat{\tau})$ and in the game $\mathcal{G}^F$ when the strategy profile is $(\sigma,\tau)$ are exactly same. 

Under this transformation, the receiver's strategy remains the same. If the sender's strategy $\sigma$ in $\mathcal{G}^F$ is a threshold strategy with a threshold $\alpha^t$ at period $t$, then $\hat{\sigma}$ is also a threshold strategy with threshold $F^{-1}(\alpha^t)$ at period $t$. So, the regular strategy profile in  $\mathcal{G}^F$ is transformed into the regular strategy profile in $\mathcal{G}^u$. Hence, the existence and unicity results in the game $\mathcal{G}^u$ can be used to derive the existence and unicity results in the game $\mathcal{G}^F$. 

\subsection*{D. Auxiliary lemmas}

For the statement and proofs in the appendix, we fix the strictly increasing continuous functions $f$ and $g$ from $I$ to $\re_+$ such that $f(0) = 0$, $g(0) = 0$.

Consider an auxiliary function $G \colon I \rightarrow \re$ is defined as $\,G(x) = x \cdot f(x) + \int_x^1 f(\theta) d\theta\,$ and recall the function $H \colon I \rightarrow \re$ defined as $H(x) = f^{-1}(\delta \cdot G(x))$.

\begin{lemma}\label{lem:func}
The following statements hold:
\begin{itemize}
\item[{[1]}] The functions $G,H$ are strictly increasing,
\item[{[2]}] The function $H$ has a unique fixed point, denoted by $\beta$,
\item[{[3]}] $H(y)>y$ for all $y < \beta$ and $H(y)<y$ for all $y > \beta$,
\item[{[4]}] $\beta\to 1$ as $\delta\to 1$ and $\beta = 1$ when $\delta = 1$.
\end{itemize}
\end{lemma}

Proof. \quad
[1] Take $0\leq x<y \leq 1$. Because $f(x)<f(y)$, we have
\begin{align*}
G(y) - G(x) &= [yf(y) + \int_y^1 f(\theta) d\theta] - [xf(x) + \int_x^1 f(\theta) d\theta] \\
& = yf(y) -xf(x) - \int_x^y f(\theta) d\theta\\
&\ge (y-x)f(y) - \int_x^y f(\theta) d\theta \, > \, 0. 
\end{align*}
Hence, $G$ is strictly increasing. The monotonicity of $H$ follows easily.

[2] We have $H(0) > 0$ and $H(1) \le 1$. As $H$ is strictly increasing, there exists $x\in I$ such that $H(x)= x$. Let $\beta=\inf\{x\in I|H(x)=x\}$. By continuity of $H$, we have $H(\beta) = \beta$. Now we will prove part 3 of the lemma. This will imply that $\beta$ is the unique solution of $H(x)=x$.

[3] We will show that the function $k(x) = f(H(x))-f(x)$ is decreasing in $x \in I$. So, take $0\leq x<y \leq 1$. We have
\begin{align*}
k(y) - k(x) & = \Big[\delta \cdot G(y) - f(y)\Big] - \Big[\delta \cdot G(x) - f(x)\Big] \\
&= \delta \cdot \Big[G(y)- G(x)\Big] - f(y) + f(x) \\
&= \delta \cdot \Big[yf(y) - xf(x) - \int_x^y f(\theta) d\theta\Big] - f(y) + f(x) \\
&= -(1-\delta y) \cdot f(y) + (1-\delta x)\cdot f(x) - \delta\cdot\int_x^y f(\theta) d\theta\\
& \le -(1-\delta y) \cdot f(x) + (1-\delta x)\cdot f(x) - \delta\cdot\int_x^y f(\theta) d\theta\\
& = \delta\cdot \big[ (y-x)\cdot f(x) - \int_x^y f(\theta) d\theta\big] \, < \, 0. 
\end{align*}
So the function $k(x) = f(H(x))-f(x)$ is indeed decreasing in $x \in I$.

Notice that $k(\beta)=0$. Assume that $y < \beta$. Then, since $k$ is decreasing, we have $k(y)>k(\beta) = 0$. Hence, $f(H(y))>f(y)$, so $H(y)>y$. Similarly, if $y >\beta$ then $H(y)<y$. 

[4] As $\beta$ is the unique fixed point of $H$, and for $\delta=1$ we have $H(1)=1$, it follows that $\beta = 1$ when $\delta = 1$. A continuity argument shows that $\beta\to 1$ as $\delta\to 1$. This completes the proof. 
\hfill \qed \vskip6pt

For the next lemma, let the game has the finite horizon $T$. Recall that $\beta^T=0$ and $\beta^t = H(\beta^{t+1})$ for all $t=1,\ldots, T-1$ and the function $V$ is defined as $V(x) = \frac{1}{x} \cdot \int_0^{x} g(\theta) d\theta$. Whenever necessary we use the notation $\beta^t(T)$ for the threshold at period $t$ to specify the horizon $T$.

\begin{lemma}\label{lem:fin}
The following statements hold:
\begin{itemize}
\item[{[1]}] $\,1 \ge
\beta> \beta^1>\beta^2>\cdots > \beta^{T}=0.$
\item[{[2]}] $\,\frac{1}{\beta^t} \cdot \int_0^{\beta^t} g^t(\theta) d\theta \, < \, \int_0^1 g^{t+1} (\theta) d\theta \,$ for $\, \delta \in [D^T,1]\,$ and $\,t=1,\ldots,T-1$
\item[{[3]}] $\beta^1(T) \rightarrow \beta$ as $T \rightarrow \infty$. More generally, $\beta^t(T) \rightarrow \beta$ as $T \rightarrow \infty$ for each $t$. 
\end{itemize}
\end{lemma}

Proof. \quad [1] By definition, $\beta^T=0$. We also have $$
\beta^{T-1} = H(\beta^T) = H(0) = f^{-1} \Bigl( \delta \cdot \int_0^1 f(\theta) d\theta \Bigr) > 0.
$$
So, $\beta^{T-1} > \beta^T$. Then inductively $\beta^{t} = H \big( \beta^{t+1} \big) > H \big( \beta^{t+2} \big) = \beta^{t+1}$ for all $t=T-2,\ldots,1$. Thus, $\beta^1>\beta^2>\cdots > \beta^{T}=0$.

As $\beta^1 > \beta^2$, we have $H(\beta^1) > H(\beta^2)=\beta^1$. So, by lemma \ref{lem:func}.3, we have $\beta^1 < \beta$. 

Finally, $\,\beta \,=\, H(\beta) \, \le \, f^{-1}(\delta f(1)) \, < \, f^{-1}(f(1)) \, = \, 1,$ so $\beta<1$. This completes the proof.

[2] It holds that
$$
\frac{1}{\beta^t} \cdot \int_0^{\beta^t} g^t(\theta) d\theta <
\frac{1}{\beta^1} \cdot \int_0^{\beta^1} g^t(\theta) d\theta \le 
\delta \cdot \int_0^1 g^t(\theta) d\theta = \int_0^1 g^{t+1} (\theta) d\theta.
$$
The first inequality follows from from substituting $a = b = 0$, $c = \beta^t$, and $d = \beta^1$ into Lemma \ref{lem:obs}. The second inequality follows from the fact that $\delta \cdot V(1) \ge V(\beta^1)$, which is true due to the assumption $\delta \in [D^T,1]$.

[3] By definition, we have $\beta^t(T) = \beta^1(T-t+1)$ for $t \le T$. So, it is sufficient to show that $\beta^1(T) \rightarrow \beta$ as $T \rightarrow \infty$. By part [1], we have $\beta^t(T) > \beta^{t+1}(T)$. So, we have $\beta^1(T-t+1) > \beta^1(T-t)$ for $t < T$. By replacing $T-t+1$ to $T$, we obtain $\beta^1(T) > \beta^1(T-1)$ for any $T > 1$.

As $H$ is strictly increasing, we have $H \big(\beta^1(T)\big) > H\big(\beta^1(T-1)\big)=\beta^1(T)$. So, by lemma \ref{lem:func}.3, $\beta^1(T) < \beta$ for all $T \in \mathbb{N}$. Hence the sequence $\big(\beta^1(T)\big)_{T \in \mathbb{N}}$ is strictly increasing and bounded above by $\beta$. Assume that the sequence converges to $y \in [0,\beta]$. We need to show that $y=\beta$. We now will assume that $y < \beta$ and show a contradiction, which will prove $y = \beta$. Denote $z =$ min $\{H(x)-x : x \in [0,y]\}$. By Lemma \ref{lem:func}.3, $H(x) > x$ for $x \le y <\beta$. By continuity of $H$, we have $z>0$.  

As the sequence $\big(\beta^1(T)\big)_{T \in \mathbb{N}}$ converges to $y$, for $\epsilon = \frac{z}{2}$, there exists $T'>0$ such that $\beta^1(T) \in [y-\epsilon,y]$ for all $T>T'$. Now for any fixed $T>T'$, consider $\beta^1(T+1) \, = \,  H\big(\beta^1(T)\big) \, \ge \, \beta^1(T) + z \, \ge \, y - \epsilon + z \, = \,y + \frac{z}{2}$. This is a contradiction as $T+1>T'$. This completes the proof. 
\hfill \qed 
\vskip6pt

Recall that the function $f:I\to \mathbb{R}$ is Lipschitz at 1 if there exist a constant $M>0$ and number $Y\in (0,1)$ such that $f(1) - f(y) \le M\cdot(1-y)$ for all $y \in [Y, 1]$.

\begin{lemma}\label{lem:infin}
The following statements hold:
\begin{itemize}
\item[{[1]}] If $f$ is Lipschitz at 1, then for every $K > 0$ there is $E\in(0,1)$ such that for all $\delta\in [E,1]$,
$$
1 - \beta \ge K \cdot (1 - \delta).
$$
\item[{[2]}] If $f$ is Lipschitz at 1, then there is $D \in [0, 1)$ such that for all $\delta \in[D,1]$,
$$
\delta\cdot V(1) \ge V(\beta).
$$
\item[{[3]}] $\,\frac{1}{\beta} \cdot \int_0^{\beta} g^t(\theta) d\theta \, < \, \int_0^1 g^{t+1} (\theta) d\theta\,$ for $\delta \in (D,1)$.
\end{itemize}
\end{lemma}

Proof. \quad
[1] The proof of part 1 is in two parts.
\vskip6pt

{\bf A.} \quad
Write $k(\delta) = 1 - \delta \cdot \beta$. We first show that
\begin{equation}\label{eqn-bd}
(1-\delta) \cdot f(1) \le \left[ f(1) - f(\beta) \right] \cdot k(\delta).
\end{equation}
Note that, by definition of $\beta$,
$$
f(\beta) = \delta \cdot G(\beta) = \delta \cdot \left[ \beta \cdot f(\beta) + \int_{\beta}^1 f(\theta) d \theta \right].
$$
This yields
$$
f(\beta) \left[ 1 - \delta \cdot \beta \right] = \delta \cdot \int_{\beta}^1 f(\theta) d \theta \le \delta \cdot (1 - \beta) \cdot f(1).
$$
Expanding brackets and adding $f(1)$ to both sides yields the inequality
$$
f(1) + f(\beta) - \delta \cdot \beta \cdot f(\beta) \le
f(1) + \delta f(1) - \delta \cdot \beta \cdot f(1).
$$
This can be rewritten into
$$
(1-\delta) \cdot f(1) \le \left[ f(1) - f(\beta \right] \cdot k(\delta).
$$
{\bf B.} \quad
We continue with the proof of the statement. Since the function $f:I\to \mathbb{R}$ is Lipschitz at 1, there exist a constant $M>0$ and number $Y\in(0,1)$ such that $f(1) - f(y) \le M\cdot(1-y)$ for all $y \in [Y, 1]$.

Take $K > 0$. Define $L = \frac{f(1)}{K \cdot M}$.
As $\delta$ tends to $1$, $k(\delta) = 1 - \delta \cdot \beta$ tends to $0$. So, there is $E\in[Y,1)$ such that $k(\delta) \le L$ for all $\delta \in[ E,1]$.
Take any $\delta \in[ E,1]$.

If $\delta=1$, then inequality (\ref{eqn-bd}) is trivially true. So suppose that $\delta\in[E,1)$. Then, using the result from A,
\begin{eqnarray*}
(1 - \delta) \cdot f(1)
& \le &
\frac{f(1) - f(\beta)}{1 - \beta} \cdot (1 - \beta) \cdot k(\delta) \cr
& \le &
M \cdot (1 - \beta) \cdot k(\delta) \cr
& \le &
M \cdot (1 - \beta) \cdot L \cr
& = &
(1 - \beta) \cdot \frac{f(1)}{K}.
\end{eqnarray*}
Since $f(1) > 0$, the part 1 of the lemma follows.  

[2] Define
$$
k(x) = g(x) - \frac{1}{x} \int_0^{x} g(\theta) d \theta
$$
for all $x\in(0,1]$. Note that $k$ is continuous and $k(1)>0$. Hence, there are $\veps > 0$ and $C \in (0, 1)$ with $k(x) \ge \veps$ for all $x \geq C$.
Define $K = \frac{V(1)}{\veps}$. Take $E\in(0,1)$ as in Lemma \ref{lem:infin}.1. In view of Lemma \ref{lem:func}.4, there is $F \in (0, 1)$ such that $\beta \geq C$ for all $\delta \in[ F,1]$.
Take $D = \max \{ E, F \}$. Take any $\delta \in [D,1]$. Then
\begin{eqnarray*}
V(1) - V(\beta)
& = &
\int_0^1 g(\theta) d \theta - \frac{1}{\beta} \int_0^{\beta} g(\theta) d \theta \cr
& \ge &
(1 - \beta) \cdot g(\beta) + \int_0^{\beta} g(\theta) d \theta - \frac{1}{\beta} \int_0^{\beta} g(\theta) d \theta \cr
& = &
(1 - \beta) \cdot \left[ g(\beta) - \frac{1}{\beta} \int_0^{\beta} g(\theta) d \theta \right] \cr
& = &
(1 - \beta) \cdot k(\beta) \cr
& \ge &
(1 - \beta) \cdot \veps \cr
& \ge &
K \cdot (1 - \delta) \cdot \veps\cr
& = & (1 - \delta) \cdot V(1).
\end{eqnarray*}
It follows that $\delta \cdot V(1) \ge V(\beta)$ for all $\delta \in [D,1]$.

[3] It holds that
$$
\frac{1}{\beta} \cdot \int_0^{\beta} g^t(\theta) d\theta <
\delta \cdot \int_0^1 g^t(\theta) d\theta = \int_0^1 g^{t+1} (\theta) d\theta.
$$
The first inequality follows the fact that $\delta \cdot V(1) \ge V(\beta)$, which is true due to the assumption $\delta \in (D,1)$.\hfill \qed \vskip6pt

\begin{lemma} \label{lem:obs} 
Let $g \colon \re \rightarrow \re$ be a non-decreasing function. Then, for any $a \le b < c \le d$,
$$
\frac{1}{c-a} \int_a^c g(x) dx \le \frac{1}{d-b} \int_b^d g(x) dx.
$$
The inequality is strict if $g$ is strictly increasing.
\end{lemma}

Proof. \quad
Define $\eta \colon [a, c] \rightarrow [b, d]$ by
$$
\eta(x) = \frac{d - b}{c - a} \cdot x + \frac{bc - ad}{c - a}.
$$
Then $\eta(a) = b$ and $\eta(c) = d$, and $\eta$ is a linear bijection between $[a, c]$ and $[b, d]$. Because $a\leq b$ and $c\leq d$ and $\eta$ is linear, we have $x\leq \eta(x)$ for all $x\in[a,c]$.

Write $h(x) = (g \circ \eta)(x)$ for all $x\in[a,c]$. Since $x\leq \eta(x)$ and $g$ is non-decreasing, $g\leq h $ on $[a, c]$. By using substitution, it follows that
\begin{eqnarray*}
\frac{1}{c-a} \int_a^c g(x) \,dx
& \le &
\frac{1}{c-a} \int_a^c h(x) \,dx \cr
& = &
\frac{1}{d-b} \int_a^c (g \circ \eta)(x) \,d \eta(x) \cr
& = &
\frac{1}{d-b} \int_b^d g(y) \,dy.
\end{eqnarray*}
This completes the proof. \hfill \qed \vskip6pt

\bibliography{sender-receiver-stopping.bib}

\providecommand{\bysame}{\leavevmode\hbox to3em{\hrulefill}\thinspace}
\providecommand{\MR}{\relax\ifhmode\unskip\space\fi MR }
\providecommand{\MRhref}[2]{%
  \href{http://www.ams.org/mathscinet-getitem?mr=#1}{#2}
}
\providecommand{\href}[2]{#2}
\begin{thebibliography}{10}

\bibitem{longcheap}
Robert~J Aumann and Sergiu Hart, \emph{Long cheap talk}, Econometrica
  \textbf{71} (2003), no.~6, 1619--1660.

\bibitem{CShumanagent}
Amos Azaria, Zinovi Rabinovich, Sarit Kraus, Claudia~V Goldman, and Ya'akov
  Gal, \emph{Strategic advice provision in repeated human-agent interactions},
  Twenty-Sixth AAAI Conference on Artificial Intelligence, 2012.

\bibitem{blume1998experimental}
Andreas Blume, Douglas~V DeJong, Yong-Gwan Kim, and Geoffrey~B Sprinkle,
  \emph{Experimental evidence on the evolution of meaning of messages in
  sender-receiver games}, The American Economic Review \textbf{88} (1998),
  no.~5, 1323--1340.

\bibitem{CrawfordSobel}
Vincent~P. Crawford and Joel Sobel, \emph{Strategic information transmission},
  Econometrica \textbf{50} (1982), no.~6, 1431--1451.

\bibitem{valuestopping}
Erik Ekstr\"{o}m and Stephane Villeneuve, \emph{On the value of optimal
  stopping games}, Ann. Appl. Probab. \textbf{16} (2006), no.~3, 1576--1596.

\bibitem{ely2017beeps}
Jeffrey~C Ely, \emph{Beeps}, American Economic Review \textbf{107} (2017),
  no.~1, 31--53.

\bibitem{ferguson1989solved}
Thomas~S. Ferguson, \emph{Who solved the secretary problem?}, Statist. Sci.
  \textbf{4} (1989), no.~3, 282--289.

\bibitem{forges1986approach}
Fran{\c{c}}oise Forges, \emph{An approach to communication equilibria},
  Econometrica: Journal of the Econometric Society (1986), 1375--1385.

\bibitem{Golosov2014}
Mikhail Golosov, Vasiliki Skreta, Aleh Tsyvinski, and Andrea Wilson,
  \emph{Dynamic strategic information transmission}, Journal of Economic Theory
  \textbf{151} (2014), 304--341.

\bibitem{GreenStokey}
Jerry~R Green and Nancy~L Stokey, \emph{A two-person game of information
  transmission}, Journal of {E}conomic {T}heory \textbf{135} (2007), no.~1,
  90--104.

\bibitem{honryo2011dynamic}
Takakazu Honryo, \emph{Dynamic persuasion}, Journal of Economic Theory
  \textbf{178} (2018), 36--58.

\bibitem{dynamics}
Simon Huttegger, Brian Skyrms, Pierre Tarr\`{e}s, and Elliott Wagner,
  \emph{Some dynamics of signaling games}, Proceedings of the National Academy
  of Sciences \textbf{111} (2014), no.~Supplement 3, 10873--10880.

\bibitem{kamenica2011bayesian}
Emir Kamenica and Matthew Gentzkow, \emph{Bayesian persuasion}, American
  Economic Review \textbf{101} (2011), no.~6, 2590--2615.

\bibitem{artconv}
Vijay Krishna and John Morgan, \emph{The art of conversation: eliciting
  information from experts through multi-stage communication}, Journal of
  Economic theory \textbf{117} (2004), no.~2, 147--179.

\bibitem{myerson1986multistage}
Roger~B Myerson, \emph{Multistage games with communication}, Econometrica:
  Journal of the Econometric Society (1986), 323--358.

\bibitem{Renault13}
J{\'e}r{\^o}me Renault, Eilon Solan, and Nicolas Vieille, \emph{Dynamic
  sender--receiver games}, Journal of Economic Theory \textbf{148} (2013),
  no.~2, 502--534.

\bibitem{renault2017optimal}
J\'{e}r\^{o}me Renault, Eilon Solan, and Nicolas Vieille, \emph{Optimal dynamic
  information provision}, Games and Economic Behavior \textbf{104} (2017),
  329--349.

\bibitem{skyrms2010flow}
Brian Skyrms, \emph{The flow of information in signaling games}, Philosophical
  Studies \textbf{147} (2010), no.~1, 155.

\bibitem{SignalsEvolution}
Brian Skyrms, \emph{Signals: Evolution, learning, and information}, Oxford
  University Press, 2010.

\bibitem{stoppingsurvey}
Eilon Solan and Nicolas Vieille, \emph{Stopping games-recent results}, Advances
  in Dynamic Games, Springer, 2005, pp.~235--245.

\end{thebibliography}

\bibliographystyle{amsplain}

\end{document}